\PassOptionsToPackage{svgnames,dvipsnames}{xcolor}

\documentclass[11pt]{preprint}

\newif\ifpreprint
\preprinttrue

\usepackage{iftex}
\usepackage{tikz}
\usetikzlibrary{backgrounds,arrows.meta,positioning,calc}
\usepackage{xspace}
\usepackage{subcaption}
\usepackage{cite}
\definecolor{linkink}{HTML}{173B65}
\definecolor{TinaCrimson}{HTML}{1C3F60}
\definecolor{CalGoldHex}{HTML}{EAF2FB}
\usepackage[colorlinks=true,allcolors=linkink,
  pdftitle={Self-Evolving Static Analysis via LLM Pseudo-Labeling},
  pdfsubject={Research preprint}]{hyperref}
\usepackage{bookmark}
\usepackage{placeins}

\titleformat{\paragraph}[runin]{\normalfont\normalsize\bfseries}{}{0pt}{#1}
\titlespacing*{\paragraph}{0pt}{0.5\baselineskip}{0.5em}
\setlist[itemize]{topsep=4pt,itemsep=3pt,parsep=0pt}
\setlist[description]{style=nextline,leftmargin=0pt,labelindent=0pt,labelwidth=0pt,labelsep=0pt,
  topsep=4pt,itemsep=5pt,parsep=0pt}

\AtBeginEnvironment{thebibliography}{\raggedright}

\usepackage{xspace}
\usepackage{xcolor}
\usepackage{listings}
\usepackage{booktabs}
\usepackage{multirow}
\usepackage{algorithm}
\usepackage{algpseudocode}
\usepackage{wrapfig}
\usepackage{subcaption}
\usepackage{fontawesome5}
\usepackage{tikz}
\usetikzlibrary{arrows.meta,positioning,shapes.geometric,fit,backgrounds}
\makeatletter
\@ifpackageloaded{tcolorbox}{}{\usepackage{tcolorbox}}
\makeatother

\newtcolorbox{takeaway}{
  colback=gray!8, colframe=gray!45, boxrule=0.4pt, arc=1pt,
  left=6pt, right=6pt, top=4pt, bottom=4pt,
  before skip=8pt, after skip=8pt,
  fontupper=\small,
}

\newtcolorbox{pilotbox}{
  colback=orange!8, colframe=orange!55!black, boxrule=0.4pt, arc=1pt,
  left=6pt, right=6pt, top=4pt, bottom=4pt,
  before skip=8pt, after skip=8pt,
  fontupper=\small,
}

\newcommand{\sysname}{SESAA\xspace}
\newcommand{\sysfull}{Self-Evolving Static Analysis Agent\xspace}

\newcommand{\msgbudget}{2000\xspace}
\newcommand{\budget}{\msgbudget}

\newcommand{\sonnet}{Claude~Sonnet~5\xspace}
\newcommand{\opus}{Claude~Opus~4.8\xspace}

\newcommand{\joern}{\textsc{Joern}\xspace}

\newcommand{\settag}[1]{\textsc{#1}\xspace}
\newcommand{\setcanonical}{\settag{S1}}   
\newcommand{\setunguided}{\settag{S2}}    
\newcommand{\setnofp}{\settag{S3}}        
\newcommand{\setnofn}{\settag{S4}}        
\newcommand{\setreal}{\settag{S5}}        

\newcommand{\fscore}{$F_1$\xspace}

\lstdefinestyle{python}{
  language=Python,
  basicstyle=\ttfamily\footnotesize,
  keywordstyle=\color{blue}\bfseries,
  commentstyle=\color{gray}\itshape,
  stringstyle=\color{teal},
  showstringspaces=false,
  breaklines=true,
  frame=single,
  numbers=left,
  numberstyle=\tiny\color{gray},
  numbersep=5pt,
}

\lstdefinestyle{scala}{
  language=Scala,
  basicstyle=\ttfamily\footnotesize,
  keywordstyle=\color{purple}\bfseries,
  commentstyle=\color{gray}\itshape,
  stringstyle=\color{teal},
  breaklines=true,
  frame=single,
  escapeinside={(*@}{@*)},
}

\definecolor{diffadd}{HTML}{1B7F3B}   
\definecolor{diffrem}{HTML}{C0392B}   
\definecolor{diffhunk}{HTML}{7F7F7F}  

\definecolor{nbnavy}{HTML}{0F3460}    
\definecolor{nblightblue}{HTML}{3F8EFC} 
\definecolor{nbteal}{HTML}{2A9D8F}    
\lstdefinelanguage{querydiff}{
  basicstyle=\ttfamily\footnotesize,
  morecomment=[f][\color{diffhunk}]{@@},
  morecomment=[f][\color{diffadd}]{+\ },
  morecomment=[f][\color{diffrem}]{-\ },
  breaklines=true,
}
\lstdefinestyle{querydiff}{language=querydiff, frame=single}

\ifpreprint
  
  \newcommand{\widefigw}{0.95\linewidth}
  \newenvironment{colfig}[1][tb]{\begin{figure}[#1]\centering}{\end{figure}}
  \newenvironment{widefig}[1][tb]{\begin{figure}[#1]\centering}{\end{figure}}
\else
  
  \newcommand{\widefigw}{\textwidth}
  
  \newenvironment{widefig}[1][t]{\begin{figure*}[#1]\centering}{\end{figure*}}
\fi

\ifpreprint
  \newenvironment{widetab}[1][tb]{\begin{table}[#1]\centering}{\end{table}}
\else
  
\fi

\usepackage{acronym}

\acrodef{LLM}{Large Language Model}
\acrodef{SAST}{Static Application Security Testing}
\acrodef{CPG}{Code Property Graph}
\acrodef{AST}{Abstract Syntax Tree}
\acrodef{CFG}{Control Flow Graph}
\acrodef{PDG}{Program Dependence Graph}
\acrodef{CWE}{Common Weakness Enumeration}
\acrodef{API}{Application Programming Interface}

\acrodef{FP}{false positive}
\acrodef{FN}{false negative}
\acrodef{TP}{true positive}
\acrodef{TN}{true negative}
\acrodef{FPR}{false positive rate}
\acrodef{FNR}{false negative rate}

\newcommand{\acReAct}{ReAct\xspace}
\newcommand{\acOWASP}{OWASP Benchmark for Python\xspace}

\newcommand{\introCVEs}{19\xspace}          
\newcommand{\introFiles}{36\xspace}         
\newcommand{\introProjects}{12\xspace}      
\newcommand{\introCaught}{10\xspace}        
\newcommand{\introCaughtPct}{53\%\xspace}   
\newcommand{\introCaughtCI}{[32\%, 73\%]\xspace}   
\newcommand{\introInRepoCaught}{9\xspace}      
\newcommand{\introInRepoTotal}{17\xspace}      
\newcommand{\introCrossRepoCaught}{0\xspace}   
\newcommand{\introRuns}{12\xspace}          

\newcommand{\codeqlf}{0.70\xspace}
\newcommand{\codeqlrecall}{0.74\xspace}

\newcommand{\codeqlfstock}{0.30\xspace}
\newcommand{\semgrepfstock}{0.19\xspace}
\newcommand{\banditfstock}{0.37\xspace}
\newcommand{\joernllmf}{0.69\xspace}

\newcommand{\sOneN}{8\xspace}
\newcommand{\sOneFmean}{0.742\xspace}
\newcommand{\sOneFsd}{0.084\xspace}
\newcommand{\sOneFciLo}{0.671\xspace}
\newcommand{\sOneFciHi}{0.812\xspace}
\newcommand{\sOnePrecMean}{0.780\xspace}

\newcommand{\sOneRecMean}{0.734\xspace}

\newcommand{\sOneBeatBaseline}{6\xspace}                    

\newcommand{\sTwoFmean}{0.713\xspace}

\newcommand{\sTwoPrecMean}{0.642\xspace}

\newcommand{\sTwoRecMean}{0.825\xspace}

\newcommand{\sOneVsTwoFp}{0.478\xspace}
\newcommand{\sOneVsTwoPrecp}{0.075\xspace}
\newcommand{\sOneVsTwoRecp}{0.127\xspace}

\newcommand{\sOneVsFourFp}{0.468\xspace}
\newcommand{\sOneVsFourPrecp}{0.245\xspace}
\newcommand{\sOneVsFourRecp}{0.304\xspace}

\newcommand{\ablFisherFstat}{7.79\xspace}
\newcommand{\ablFisherFp}{0.254\xspace}
\newcommand{\ablFisherPrecStat}{10.58\xspace}
\newcommand{\ablFisherPrecp}{0.102\xspace}

\newcommand{\ablBayesTwoPostD}{-0.31\xspace}   
\newcommand{\ablBayesTwoPostSd}{0.50\xspace}
\newcommand{\ablBayesTwoProb}{73\%\xspace}     
\newcommand{\ablBayesThreeD}{-1.05\xspace}
\newcommand{\ablBayesThreePostD}{-0.77\xspace}
\newcommand{\ablBayesThreePostSd}{0.52\xspace}
\newcommand{\ablBayesThreeProb}{93\%\xspace}

\newcommand{\ablBayesFourPostD}{-0.34\xspace}
\newcommand{\ablBayesFourPostSd}{0.50\xspace}
\newcommand{\ablBayesFourProb}{75\%\xspace}

\newcommand{\costOneAgentIn}{254M\xspace}
\newcommand{\costOneAgentOut}{490K\xspace}
\newcommand{\costOneCriticIn}{1.68M\xspace}    
\newcommand{\costOneCriticOut}{213K\xspace}   

\newcommand{\sThreeVsOneFp}{0.091\xspace}
\newcommand{\sThreeVsOneJp}{0.155\xspace}
\newcommand{\sThreeVsOnePrecp}{0.274\xspace}
\newcommand{\sThreeVsOneRecp}{0.244\xspace}

\newcommand{\tkMultiCalibGapMean}{0.19\xspace}
\newcommand{\tkMultiCalibGapSd}{0.07\xspace}
\newcommand{\tkMultiCalibGapMin}{0.11\xspace}
\newcommand{\tkMultiCalibGapMax}{0.26\xspace}

\newcommand{\tkMultiN}{8\xspace}
\newcommand{\tkMultiTickets}{2158\xspace}
\newcommand{\tkMultiFindings}{1108\xspace}
\newcommand{\tkMultiSolved}{665\xspace}
\newcommand{\tkMultiSolvedFP}{568\xspace}
\newcommand{\tkMultiSolvedReal}{97\xspace}
\newcommand{\tkMultiPrecisionPooled}{85\%\xspace}    
\newcommand{\tkMultiPrecisionMean}{89\%\xspace}       
\newcommand{\tkMultiPrecisionSd}{0.11\xspace}
\newcommand{\tkMultiPrecisionMin}{68\%\xspace}
\newcommand{\tkMultiPrecisionMax}{100\%\xspace}

\newcommand{\tkMultiTargetSig}{7\xspace}                   
\newcommand{\tkMultiSelectSig}{8\xspace}                   
\newcommand{\tkMultiCalibSig}{4\xspace}                    
\newcommand{\tkMultiCalibPos}{8\xspace}                    
\newcommand{\tkMultiTargetP}{1.1\times10^{-35}}   
\newcommand{\tkMultiSelectP}{5.2\times10^{-71}}   

\newcommand{\tkMultiFNSig}{8\xspace}                       
\newcommand{\tkMultiFNn}{159\xspace}                       
\newcommand{\tkMultiFNprec}{61\%\xspace}                   
\newcommand{\tkMultiFNPrecMean}{70\%\xspace}                
\newcommand{\tkMultiFNPrecSd}{17\%\xspace}                  
\newcommand{\hToolCalls}{711\xspace}        

\newcommand{\hEditQuery}{85\xspace}
\newcommand{\hWriteQuery}{37\xspace}

\newcommand{\corpustotal}{1{,}230\xspace}
\newcommand{\corpusvuln}{452\xspace}
\newcommand{\corpussafe}{778\xspace}
\newcommand{\corpusclasses}{14\xspace}

\title{Self-Evolving Static Analysis via LLM Pseudo-Labeling}

\author{%
  \begin{tabular}{@{}c@{}}
    Rui Melo\textsuperscript{1,2}\quad
    Andre Duarte\textsuperscript{1,*}\quad
    Aditya Oke\textsuperscript{1,*}\quad
    Shubham Gandhi\textsuperscript{1}\\[4pt]
    Nachiket Kotalwar\textsuperscript{1}\quad
    Rui Abreu\textsuperscript{2}\quad
    Lei Li\textsuperscript{1}\quad
    Carolyn Rose\textsuperscript{1}\\[8pt]
    \small\textsuperscript{1}Carnegie Mellon University\\
    \small\textsuperscript{2}FEUP
  \end{tabular}%
}
\date{}

\begin{document}
\maketitle

\begingroup
\renewcommand{\thefootnote}{\fnsymbol{footnote}}
\footnotetext[1]{Equal second authorship.}
\endgroup

\begin{abstract}

Security bugs often slip past static analysis not because they are impossible to detect, but because detection rules have not yet been written for the code in front of us. New frameworks, APIs, and coding idioms continually create such gaps, leaving developers to discover failures and manually encode new rules.

We present \sysname{}, a framework in which an \ac{LLM} agent starts from no rules and iteratively evolves a set of readable and auditable \joern{} queries. The agent receives feedback only from a second LLM critic, which inspects flagged and unflagged code to identify likely false alarms and suspected misses, without access to ground-truth labels. The agent then uses this feedback to revise its queries.

On the OWASP Benchmark for Python, \sysname{} reaches a mean \fscore{} of $\sOneFmean{}$ across $\sOneN{}$ independent runs, well above the default configurations of five widely used tools and comparable to the strongest of them once it is tuned with access to ground-truth labels that \sysname{} never sees. 
In a pilot on real-world Python projects, queries induced from code available before a vulnerability was introduced detect \introCaught{} of \introCVEs{} subsequent CVEs, including \introInRepoCaught{} detected by rules written for the same repository.

The resulting detector is a set of executable, human-readable queries that engineers can inspect, modify, and run independently. These results suggest that LLM agents can turn static-analysis rule maintenance from a manually authored process into an iterative process of automated discovery and repair.

\end{abstract}

\section{Introduction}
\label{sec:introduction}

\begin{widefig}[t]
  \begin{subfigure}[t]{0.49\linewidth}
    \caption{Vulnerable (\texttt{BenchmarkTest00099})}
    \label{fig:motiv-vuln}
    \begin{lstlisting}[style=python,numbers=none,basicstyle=\ttfamily\fontsize{6.8}{8.4}\selectfont]
# tainted
param = request.form.get("...")  
# ...param flows into bar...
bar = conf.get("section", "keyB")
sql = f"SELECT ... WHERE password='{bar}'"
# tainted sink
cur.execute(sql)
    \end{lstlisting}
  \end{subfigure}
  \hfill
  \begin{subfigure}[t]{0.49\linewidth}
    \caption{Safe (\texttt{BenchmarkTest00286})}
    \label{fig:motiv-safe}
    \begin{lstlisting}[style=python,numbers=none,basicstyle=\ttfamily\fontsize{6.8}{8.4}\selectfont]
# ...same shape as (a)...
param = wrapped.get_form_parameter("...")
bar = conf.get("section", "keyA")
# parameterized
sql = "SELECT ... WHERE password=?" 
# bound arg
cur.execute(sql, (bar,))
    \end{lstlisting}
  \end{subfigure}

  \vspace{2pt}

  \begin{subfigure}[t]{\linewidth}
    \caption{A taint-aware detector refinement that tells (a) and (b) apart (illustrative)}
    \label{fig:motiv-diff}
    \begin{lstlisting}[style=querydiff,numbers=none]
// ANY argument tainted -> false-positive on (b)
- cpg.call.name("execute|executemany|executescript")
-   .filter(c => c.argument                         
-     .reachableByFlows(SOURCES).nonEmpty)   
// only the SQL-string arg -> spares parameterized (b)
+ cpg.call.name("execute|executescript")
+   .filter(c => c.argument.argumentIndex(1)        
+     .reachableByFlows(SOURCES).nonEmpty)        
    \end{lstlisting}
  \end{subfigure}
  \caption{\textbf{Why detectors must evolve.} Two near-identical OWASP cases differ only in how user input reaches the database: (a) concatenates it into the SQL string (vulnerable), while (b) binds it as a parameter (safe). 
  A detector that flags any \texttt{execute()} call with nearby tainted input fires on both --- a false positive on (b). 
  Telling them apart requires understanding the mechanism behind SQL injection and taint analysis.
  The user data must reach the \emph{query-string} argument (argument~1), as sketched in (c).
  Production-ready queries have to provide this distinction.
  However, keeping such rules correct as frameworks, APIs, and idioms change is a maintenance burden.
  }
  \label{fig:motivating}
\end{widefig}

Static-analysis tools~\cite{yamaguchi2014, semgrep, bandit, codeql}
are widely used in software pipelines
allowing for scanning source code for security bugs.
Security experts write these tools' queries by hand 
and must revise them as frameworks and codebases change.
Each revision requires the expert to spot a missed vulnerability or a \ac{FP}.
It needs to diagnose why the rule failed, and rewrite it without breaking what already works (Figure~\ref{fig:motivating} shows one such repair).
This process is rather slow and depends heavily on the expert.
A narrow rule misses new patterns, and a broad one produces the \acp{FP} that have long held back adoption~\cite{johnson2013why,christakis2016what,bessey2010few,sadowski2018lessons}.

Recent work has started to hand this labour to \acp{LLM}, asking a model to read the code and report bugs directly~\cite{sun2024llm4vuln,10.1145/3786165.3788441} or to write rules from external inputs such as published CVE reports or fixing patches~\cite{qlcoder2025,irsan2026querygen,mocq2025}.
The first kind leaves little that can be reused in later scans, while the second ties its findings to previously documented vulnerabilities.
We ask whether an \ac{LLM} agent can, instead, maintain its own rules by observing where a rule fails, diagnosing why, and rewriting it, as a security engineer would.

We present \sysname{}, an \ac{LLM} agent that starts with no rules and builds a set of queries for the open-source analyzer \joern{}, which represents code as a searchable graph (Figure~\ref{fig:framework}). The agent operates in rounds, editing its queries and running them over a body of code that no one has marked as vulnerable or safe.

A second \ac{LLM}, the critic, reviews what the queries flagged, judges each finding as a likely bug or a false alarm, and points out places where it suspects a bug was missed.
We call these judgments pseudo-labels, because they stand in for labels that no one supplies.
The run keeps the set of queries that scores best on them, and no finding is ever checked against ground-truth labels.
The result is an ordinary set of queries that an engineer can read, edit, and run without the agent (\S\ref{sec:system}).

\paragraph{\textbf{Contributions.}}
Our contributions are as follows.
(1) \sysname{}, a framework where an agent develops and maintains detection queries by evolving them from a critic's pseudo-labels, without ground-truth labels.
(2) Evolved detectors that reach a mean \fscore{} of $\sOneFmean{}$ over $\sOneN{}$ independent runs on the OWASP Benchmark for Python~\cite{owaspbenchmarkpy}, well above the default configurations of five widely used tools and comparable to CodeQL tuned with access to the labels.
(3) A pilot study on real-world Python projects, in which queries written from code taken before a vulnerability was introduced flagged \introCaught{} of \introCVEs{} later CVEs.

The remainder of the paper describes the system and its evolutionary loop (\S\ref{sec:system}), details the evaluation setup (\S\ref{sec:methodology}), reports results (\S\ref{sec:results}), reviews related work (\S\ref{sec:background}), discusses the mechanism and threats to validity (\S\ref{sec:discussion}, \S\ref{sec:threats}), and concludes (\S\ref{sec:conclusion}).

\section{System Architecture}
\label{sec:system}

\sysfull{} (\sysname{}), pairs two roles, shown in Figure~\ref{fig:framework}.
The agent writes and repairs detection queries, and the critic, a second \ac{LLM}~\cite{zheng2023judge}, reviews what those queries report.
Because no labels exist to say whether they are right, the critic judges each finding as a likely true or false positive and turns its doubts into \emph{tickets}, concrete items for the agent to resolve, either a false alarm to remove or a suspected miss to close. 
Starting from an empty set of queries, the agent repeatedly commits a batch of edits, which we call a \emph{mutation}, and one committed mutation together with the re-run that scores it makes up a \emph{generation}.
After a fixed budget of generations, the run returns the best-scoring suite (Algorithm~\ref{alg:evolution_loop}).
A security team can read, review, and run the suite independently of the agent, since it consists of ordinary code and has no dependency on an LLM.

\begin{widefig}[t]
  \includegraphics[width=\widefigw]{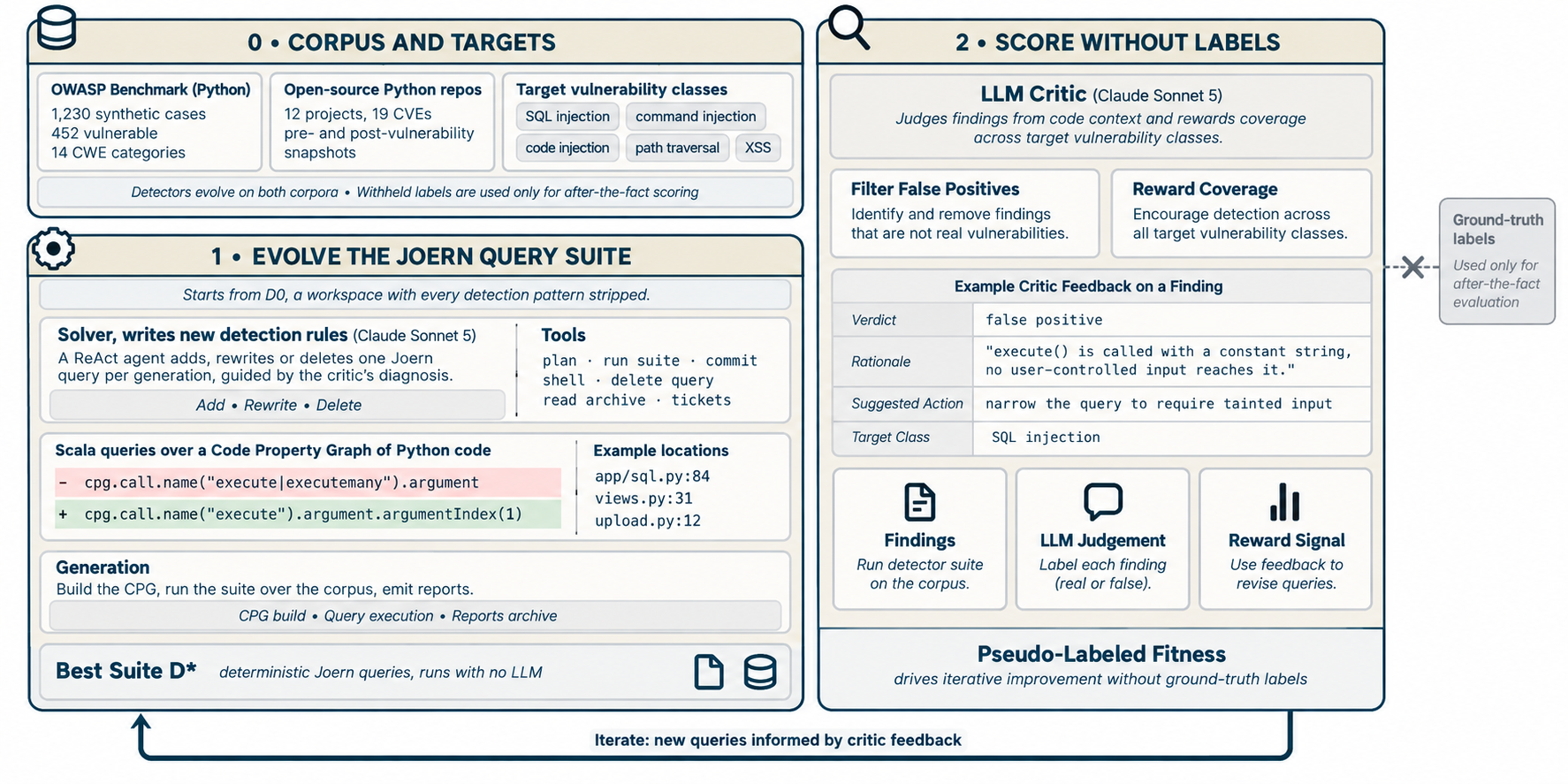}
  \caption{\textbf{Overview of \sysname{}.}
  The \acReAct{} agent repeatedly mutates the \joern{} queries and re-runs them over the corpus.
  A critic pseudo-labels the reports, estimating precision from every report and rewarding coverage across vulnerability classes.
  Ground-truth labels are never accessible.
  Over time, queries that receive stronger assessments from the critic are preferentially retained and propagated.
  }
  \label{fig:framework}
\end{widefig}

\textbf{Code Property Graph and Engine.}
\label{sec:cpg}
We build on \joern{}~\cite{yamaguchi2014}. Its \ac{CPG} merges syntax (\ac{AST}), execution paths (\ac{CFG}), and data and control dependencies (\ac{PDG}) in one graph, so a single query can express the taint flows that injection vulnerabilities need.
Every query is ordinary Scala, a general-purpose language the agent can edit directly.
Additionally, it is open source with Python frontend, so the loop can run it repeatedly over the corpus.

\textbf{Evolutionary Algorithm.}
\label{sec:algorithm}
The agent is a \acReAct{} agent~\cite{yao2022react}, an \ac{LLM} that alternates between reasoning and calling tools, and in each generation it acts on the critic's diagnosis of where the current queries fail, which lists its tickets and coverage gaps (\S\ref{sec:fitness}).
The difficulty is that every repair risks damage elsewhere, the stability--plasticity dilemma~\cite{mermillod2013stability}, since a query tightened to remove one false alarm may stop catching a real bug.
Two things limit that risk.
The run keeps the highest-fitness suite seen so far, so a bad mutation costs one generation and not the run, and the fitness in \S\ref{sec:fitness} penalizes losing either precision or coverage.
The repairs that hold up tend to capture \emph{why} a flagged case is safe (i.e., it uses escaped input or a parameterized binding) rather than just narrowing which cases match, and the agent often records that rationale inline as a comment.
Algorithm~\ref{alg:evolution_loop} gives the procedure.

The agent has seven tools, which fall into three groups.
To act, it has a shell for running commands, a planning tool for deliberating between actions, and a query-deletion tool, and it ends each generation with a commit tool.
To check its work, it has a corpus-evaluation tool that runs the current suite and returns the critic's scored diagnosis, and a ticket tool for listing, reviewing, or dismissing false-positive and false-negative tickets.
To consult its own history, it has an archive-inspection tool that reads prior generations' suites.

\begin{algorithm}[t]
\caption{\sysname{} Evolution Loop}
\label{alg:evolution_loop}
\begin{algorithmic}[1]
\Require Corpus $C$;\ baseline detectors $D_0$ (all patterns stripped);\ message budget $B$
\Require Agent $\mathcal{M}_{\pi}$: a \acReAct{} agent under model $\pi$ that reads the critic's diagnosis and applies one \emph{mutation} to the detector suite --- writing, editing, or deleting a \joern{} query
\Require Critic $\mathcal{K}\!:(D(C),\, C,\, \mathcal{V}) \mapsto (\hat{p},\, \hat{c},\, \phi,\, \mathcal{V}')$, scoring reports without labels and emitting a diagnosis $\phi$ (FP tickets, FN tickets, coverage gaps)
\Require Verdict store $\mathcal{V}$: persistent per-report verdicts and tickets (initially $\varnothing$)
\Ensure  Best detector suite $D^{\star}$

\State $D \gets D_0$;\quad $\mathcal{V} \gets \varnothing$;\quad $\phi \gets \varnothing$;\quad $F^{\star} \gets -\infty$
\State \textbf{The budget $B$ bounds the loop:} each iteration consumes messages, so $B$
       sets how many mutate--evaluate cycles (generations) the run can perform.
\While{messages remaining under $B$}
    \State \textbf{Mutate:} the agent applies one mutation, guided by the diagnosis
    \State $D \gets \mathcal{M}_{\pi}\big(D,\, C,\, \phi,\, \mathcal{V}\big)$

    \State \textbf{Evaluate (one generation):} build the \acs{CPG} and run $D$ over $C$, yielding reports $D(C)$

    \State \textbf{Critic:} per-report verdicts against the carried store; no ground truth
    \State $(\hat{p},\, \hat{c},\, \phi,\, \mathcal{V}) \gets \mathcal{K}\big(D(C),\, C,\, \mathcal{V}\big)$
    \State $a \gets a_{\max}\cdot\min\!\big(1,\ |D(C)|/n^{\ast}\big)$;\quad
           $F \gets \min\!\big(1,\ \sqrt{\hat{p}\hat{c}}\,(1+a)\big)$

    \State \textbf{Keep best:} retain the highest-fitness suite seen so far
    \If{$F > F^{\star}$}
        \State $D^{\star} \gets D$;\quad $F^{\star} \gets F$
    \EndIf
\EndWhile

\State \Return $D^{\star}$
\end{algorithmic}
\end{algorithm}

\subsection{Pseudo-Labeled Fitness}
\label{sec:fitness}

The fitness must reward both precision and coverage without ground-truth labels.
We call it \emph{pseudo-labeled}: the critic judges each finding as a likely \ac{TP} or \ac{FP} from the surrounding code and context, and the suite's fitness is aggregated from those judgments.
The judgments are imperfect, since \acp{LLM} remain unreliable at vulnerability reasoning~\cite{ding2024howfararewe}, so we measure the critic's accuracy against the withheld labels (\S\ref{sec:tickets}).
The signal is therefore weak supervision from a pretrained model, not supervision-free.

We use these judgments to capture two complementary properties of a query suite: precision, through the critic's assessment of individual findings, and coverage, through the vulnerability classes represented among those findings.

We formalize this process as follows.
Let $C$ be the corpus of applications and $Q = \{q_i\}_{i=1}^{n}$ a suite of queries.
Each query maps the corpus to a set of \emph{reports}, where a report is one alert: a code location the query flags as a candidate vulnerability of a given type.
Write $Q(C)$ for the pooled reports of the suite.
The fitness $F(Q)$ is computed from $Q(C)$ and the critic's verdicts and tickets on it, never from ground-truth labels.

\textbf{Estimated precision.}
The critic assigns every report in $Q(C)$ a verdict, \ac{TP} or \ac{FP}, stored in a persistent knowledge base keyed by the report's identity and reused in later generations, so findings that survive a mutation are never re-evaluated.
To prevent high-volume queries from dominating the score, findings are grouped by \text{query} and \text{vulnerability type}. 
For each group ($g$), estimated precision is the fraction of findings judged as \ac{TP}, $
\hat{p}_g := \frac{\left|\{r \in g : v(r)=\textsf{TP}\}\right|}{|g|}
$,
where $v(r)$ is the stored verdict for report $r$. The suite-level estimated precision is the unweighted mean across groups, $\hat{p}:=\frac{1}{|G|}\sum_{g\in G}\hat{p}_g.$

\textbf{False-positive tickets.} Reports the critic marks as \acp{FP} are surfaced to the agent as \emph{tickets}: a persistent, per-report list carried across generations.
The agent clears a ticket by tightening the query so the report is no longer emitted, or by \emph{contesting} the verdict with an argument that the report is a \ac{TP}; a contest re-invokes the critic on that single report, and the verdict flips only if the critic is persuaded.

\textbf{Type coverage.}
Recall is unobservable without labels, so the fitness needs a proxy for it.
We measure how many distinct vulnerability \emph{classes} (SQL injection, path traversal, and so on) the suite exercises: $\hat{c}_{\text{type}} = \frac{\lvert\{\, t \in T : \exists\, r \in Q(C),\; \tau(r) = t \,\}\rvert}{\lvert T \rvert}$, where $T$ is the target taxonomy and $\tau(r)$ a report's type. This rewards a suite for covering the classes it targets, not for flagging as many files as possible, and is indifferent to how vulnerable the corpus happens to be. For the OWASP evaluation, $T$ comprises the benchmark's own \corpusclasses{} CWE categories. The agent is never told which files are vulnerable, but it is told which classes count toward coverage. Its queries may use a broader 24-type vocabulary recognized by the harness, including types outside the OWASP taxonomy (e.g., SSRF and IDOR); unrecognized labels score as unknown with precision $0$.
Type coverage measures breadth rather than depth. A class is covered after a single report, so $\hat{c}_{\text{type}}$ saturates once all target classes are touched, regardless of remaining false negatives. It therefore provides a useful cold-start signal but no gradient for missed instances; the FN tickets below provide that signal.

\textbf{False-negative tickets.}
\ac{FP} tickets let the critic audit reported findings; \ac{FN} tickets audit suspected misses, necessarily suspected since the critic never sees withheld labels (\S\ref{sec:ticket-method} reports how often these suspicions are confirmed). In each generation, the critic reviews a limited corpus subset, one vulnerability class at a time, and opens an \ac{FN} ticket for each vulnerable-looking operation lacking a report. The agent resolves it by broadening or adding a query, or by contesting its vulnerability; the critic dismisses the ticket only if persuaded.

Let $\mathcal{R}$ be the \ac{FN} tickets opened over the reviewed portion so far, partitioned into resolved and open. Coverage is
$\hat{c} = \frac{\lvert \mathcal{R}_{\text{resolved}} \rvert}{\lvert \mathcal{R}_{\text{resolved}} \rvert + \lvert \mathcal{R}_{\text{open}} \rvert}$,
the fraction of critic-identified misses closed so far --- a bounded $[0,1]$ recall proxy that rewards each additional recovered instance. This stays free of ground-truth labels because the denominator ranges only over what the critic has actually reviewed, and the critic, not the agent, controls which portions are reviewed, so coverage can only rise by closing tickets, i.e.\ by recovering missed vulnerabilities.

\textbf{Fitness Proxy.}
Fitness is the geometric mean of the two proxies, corrected by a single bounded activity term and clamped to $[0,1]$: $ F(Q) \;=\; \min\!\Big(1,\ \sqrt{\hat p \cdot \hat c}\,\bigl(1 + a\bigr)\Big)$, where $a \;=\; a_{\max}\cdot\min\!\Big(1,\ \frac{|Q(C)|}{n^{\ast}}\Big) $,
with ceiling $a_{\max}=0.05$ and saturation count $n^{\ast}=10$ reports.
We use the geometric mean, a standard balanced score under class imbalance~\cite{kubat1997}.
$a$ ramps from $0$ to its $5\%$ ceiling over a rule's first ten reports and stays there, a tie-breaker rewarding output without letting an active-but-worse suite outrank a better one. Both constants are chosen conservatively, not tuned.

\section{Illustrative Example}
\label{sec:example}

This example follows one early repair through the loop of \S\ref{sec:system}.
Like the SQL case in Figure~\ref{fig:motivating}, the XPath-injection query relied on a coarse taint rule that could not tell why a construct was safe, so it raised \acp{FP} on benign code.

\noindent\textbf{The critic doubts a finding.}
After a generation, the critic reviews what the XPath query flagged.
It never sees a label, so it judges each finding from the surrounding code and opens a \ac{FP} ticket for the findings that look benign, here calls that pass their values as bound parameters or escape them first.

\noindent\textbf{The agent repairs the query.}
The agent reads the ticketed cases and the code around them, identifies two safe idioms, and adds an exclusion for each to the query (Figure~\ref{fig:xpath-diff}).
The query now ignores calls whose query literal contains an escape entity and calls that use a \texttt{name=} binding.
The edit captures \emph{why} these cases are safe, and the agent records that rationale in an inline comment.

\begin{figure}[hbt!]
\begin{lstlisting}[style=querydiff,numbers=none]
  cpg.call.name("xpath|select")
+   // exclude XML-escaped queries (value neutralised)
+   .filterNot(c => c.method.ast.isLiteral
+     .code(".*&apos;.*|.*&quot;.*").nonEmpty)
+   // exclude parameterized binding: xpath(q, name=...)
+   .filterNot(c => c.code.matches(".*\bname\s*=.*"))
    .filter(c => c.argument
      .reachableByFlows(SOURCES).nonEmpty)
\end{lstlisting}
\caption{\textbf{Semantic repair to the XPath-injection query.} The agent's edit (green) adds two exclusions to the original rule (black): XML-escaped arguments and calls that pass values via a named parameter (\texttt{name=...}) rather than concatenation.}
\label{fig:xpath-diff}
\end{figure}

\noindent\textbf{The suite is scored again.}
The agent commits the mutation, and the corpus-evaluation tool re-runs every query over the code property graph and rescores the findings (Algorithm~\ref{alg:evolution_loop}).
With the escaped and parameterized cases excluded, the XPath group's estimated precision rises, the fitness improves over the best suite so far, and the run keeps the new suite (\S\ref{sec:fitness}).

\noindent\textbf{What the example shows.}
The repair is small, and it is the kind an engineer would make.
It separates safe constructs from unsafe ones and does not suppress the query, and the rationale comments make the change reviewable without the agent.
The exchange can also run the other way, as in the ticket of \S\ref{sec:tickets} where the agent contests the critic's verdict and the critic reverses it, so the signal that drives the repair is checked from both sides.

\section{Methodology}
\label{sec:methodology}

This section describes our experimental setup: the evaluation corpus (\S\ref{sec:corpus}), the baseline and agent configuration (\S\ref{sec:config}), the models we use (\S\ref{sec:models}), the experimental settings (\S\ref{sec:settings}), the metrics we use to evaluate the evolved detectors (\S\ref{sec:metrics}), the protocol for analyzing the ticket mechanism (\S\ref{sec:ticket-method}), the statistical tests (\S\ref{sec:statistical-tests}), and the introduction-time detection dataset (\S\ref{sec:intro-dataset}).

\subsection{Evaluation Corpus}
\label{sec:corpus}

We evaluate \sysname{} on the \acOWASP{}~\cite{owaspbenchmarkpy}, a synthetic suite where each file contains exactly one test case labeled vulnerable or safe.
It spans \corpusclasses{} vulnerability categories and \corpustotal{} test cases in total, \corpusvuln{} vulnerable and \corpussafe{} safe.
We evaluate one language, Python, with \sOneN{} \setcanonical{} seeds and three ablations.

We extend the evaluation to a small corpus of real-world Python projects, each containing at least one known vulnerability. Each repository appears in two states: a pre-vulnerability snapshot, against which the agent writes its own queries, and a post-vulnerability snapshot, against which we check whether those queries flag the introduced vulnerability.

\subsection{Evolution Configuration}
\label{sec:config}

We report a single configuration of \sysname{}'s loop. We choose a \sonnet{}~\cite{anthropic-sonnet5} agent paired with an \opus{}~\cite{anthropic-opus48} critic (\S\ref{sec:models}), evolving detectors on the \acOWASP{}, orchestrated with the Inspect AI framework~\cite{inspect-ai}.
The loop starts from a workspace with all detection patterns stripped, forcing the agent to discover detectors from scratch rather than tuning an existing suite.

\textbf{Generations and mutations.}
A \emph{mutation} is the change the agent commits to the detector suite by writing, editing, or deleting \joern{} queries.
A \emph{generation} is one committed mutation, after which the suite is re-run and scored.
Before committing, the agent may edit, measure the result, and read the critic's feedback as many times as it likes.

\textbf{Budget and determinism.}
Each generation's react loop is bounded by a budget of \budget{} messages, stated verbatim in the system prompt, after which the workspace is auto-committed even if the agent never explicitly commits.
Ground-truth labels are never placed in the agent's context or tool results. We compute the true \fscore{} separately, purely for the evaluation of \S\ref{sec:results}. We write the critic's per-finding verdicts once into a persistent store and reuse them whenever a finding survives a mutation unchanged, so re-evaluating an untouched detector does not re-invoke the judge.

\textbf{Baselines.}
We run a panel of widely used static and dynamic-analysis tools, CodeQL~\cite{codeql}, Semgrep~\cite{semgrep}, Bandit~\cite{bandit}, Bearer~\cite{bearer}, and OWASP ZAP~\cite{owaspzap}, each in two configurations, scored with the official OWASP scorecard. The out-of-the-box configuration runs the community queries. CodeQL uses the \emph{python-security-extended} suite with the GitHub Security Lab pack, and Semgrep uses the community packs \emph{p/python}, \emph{p/flask}, \emph{p/secrets}, \emph{p/security-audit}, and \emph{p/owasp-top-ten}.
The tuned configuration upgrades those community queries by having an LLM agent evolve them against the withheld labels. We run this tuning with three backbones (\sonnet{}, Kimi-K2.5, GLM-5) at three window sizes (30, 60, 100 retained messages), for 9 tuning runs per tool. We report the best-performing queries from each tool's 9-run panel, scored against the withheld OWASP labels after tuning.
Unlike \sysname{}, which never observes the withheld labels at any point, the best-of-9 selection for every baseline tool is made \emph{with} access to them.
\sysname{} matching or exceeding a baseline tuned this way is therefore evidence under a harder information constraint, not a like-for-like accuracy race (\S\ref{sec:baselines}).

We run the same 3-backbone $\times$ 3-window panel with \joern{} itself as a sixth baseline. 
A single agent turn writes queries directly against the withheld labels, with no critic, tickets, or further generations, and we report the best of the 9 configurations. \joern{} ships no built-in ruleset, so an empty out-of-the-box query set would trivially score zero rather than a weaker-but-nonzero point.

\subsection{Models}
\label{sec:models}

The task demands a backbone fluent in both Python and Scala. We use \sonnet{}~\cite{anthropic-sonnet5} as the agent, chosen for its strong code-generation and agentic 
performance~\cite{mbpp,humaneval,swebench}, and the more capable \opus{}~\cite{anthropic-opus48} as the critic, mirroring a realistic deployment where a cheaper model handles high-volume exploration while a more capable model adjudicates the smaller set of findings it surfaces.

\subsection{Experimental Settings}
\label{sec:settings}

We organize the evaluation into five settings, identified throughout as \setcanonical{}--\setreal{} and summarized visually in Figure~\ref{fig:combined-results}.
\setcanonical{} is the configuration of \S\ref{sec:config}, repeated across \sOneN{} independent runs to quantify the variance of a stochastic loop (\S\ref{sec:trajectory}). \setunguided{} to \setnofn{} are ablations, each run with five independent seeds, that remove one component at a time while holding the corpus, message budget, tooling, and model pair fixed, so the difference in ground-truth quality measures the contribution of the removed component. \setreal{} moves the loop to a corpus of real-world Python projects.

In \setunguided{}, the agent keeps its budget, tools, and corpus and still sees the raw findings of its queries, but receives no critic verdicts, tickets, or fitness. The returned suite is chosen exactly as in \setcanonical{}, so \setunguided{} measures what an agent achieves by writing and running detectors alone. \setnofp{} keeps the critic's precision estimate but removes the itemized \ac{FP} tickets and contest step of \S\ref{sec:fitness}, isolating the ticket surface's contribution from that of the score it feeds. \setnofn{} removes the \ac{FN} tickets, so coverage reverts to the presence-based $\hat{c}_{\text{type}}$, which changes both the ticket surface and the recall term of the fitness.

\setreal{} keeps the critic and the tickets and changes the corpus and the fitness proxy.
Pre-introduction snapshots contain no vulnerabilities, so precision and recall against labeled findings are undefined and the proactive fitness of \S\ref{sec:intro-dataset} replaces them.
\setreal{} therefore tests whether an agent that reasons over clean code writes queries that catch vulnerabilities introduced later, not whether the fitness design of \S\ref{sec:fitness} holds on real code.

\subsection{Evaluation Metrics}
\label{sec:metrics}

We report precision, recall, \fscore{}, and Youden's J computed per finding against the withheld labels.
These labels are assigned a priori and are never visible to the agent.
Precision is the share of a suite's alerts that point at a real vulnerability, so a precision of 0.8 means four of every five alerts are real.
Recall is the share of the benchmark's vulnerable test cases that the suite flags.
The \fscore{} is their harmonic mean and is high only when both are high.
Youden's $J$ is the true-positive rate minus the false-positive rate, so a suite that flags every file scores zero.
The benchmark has \corpusvuln{} vulnerable and \corpussafe{} safe test cases, so a suite that flags nothing has zero recall and one that flags everything has a false-positive rate of one.
For the introduction-time benchmark (\S\ref{sec:intro-dataset}), we report whether the suite detects the vulnerability at the commit that introduces it, and whether the ticket mechanism then corrects any resulting \acp{FP}.

\subsection{Analyzing the Ticket Mechanism}
\label{sec:ticket-method}

One of \sysname{}'s central components is the ticket mechanism, which lets the critic flag a finding and the agent repair or contest it.
We fix the protocol for one of two separate questions: whether the mechanism \emph{behaves as designed}, meaning the critic's judgment tracks ground truth and the agent acts on it.
The other question, whether \emph{having} the mechanism \emph{improves the detector}, is answered separately by the \setnofp{} ablation (defined in \S\ref{sec:settings}; results in \S\ref{sec:trajectory}), and by the \setnofn{} ablation for \ac{FN} tickets, which compare outcomes with the ticket surface on versus off across independent seeds.

\paragraph{\textbf{Labeling tickets against ground truth.}}
Checking whether a ticket is accurate requires knowing whether the finding behind it is a genuine \ac{FP}, so we check every opened ticket and every repair made against the withheld labels. The critic opens a ticket for each finding it judges a \ac{FP}. The agent may then repair the query or contest the ticket, and the critic accepts or rejects the contest. We label each ticketed finding as a genuine \ac{FP} or a real vulnerability, judging both whether the critic's ticket (and contest verdict) is correct, and whether the agent's repair removed a genuine \ac{FP} or destroyed a real detection.

\subsection{Statistical Tests}
\label{sec:statistical-tests}

The ticket mechanism setup raises three questions. First, whether a \ac{FP} ticket actually points at a genuine \ac{FP}. Second, whether repairs concentrate on the tickets the critic flagged rather than falling just as likely on anything else. Third, whether the critic's precision estimate is systematically biased relative to the withheld ground truth. 
For each, we compute the null distribution by exact enumeration rather than a normal approximation. 
We compute Fisher's exact test~\cite{fisher1922interpretation} for the first question (whether \ac{FP} tickets are independent of ground-truth \ac{FP} status), a one-sided hypergeometric test for the second (random draw versus selective response), and an exact Wilcoxon signed-rank test~\cite{wilcoxon1945individual,lehmann1975nonparametrics} for the third.

We perform two additional checks of the mechanism's overall behavior. First, we test whether the critic's fitness score tracks the withheld ground truth over the full run, using Pearson correlation with a $t$-test on the coefficient~\cite{student1908probable} and Spearman correlation to assess rank agreement. Second, we test whether the critic's judgment extends to the recall side by comparing the fraction of \ac{FN} tickets that correspond to real vulnerabilities against the per-class base rate of a random file, using a one-sided Fisher exact test~\cite{fisher1922interpretation}.

For ablation comparisons (\S\ref{sec:settings}), Welch's $t$-test compares each ablation's mean F1 against \setcanonical{}, the reported configuration repeated across seeds (\S\ref{sec:settings}). 
For the ticket mechanism, we compute each test above per-seed, then combine across seeds via Fisher's method (a $\chi^2$-test on pooled $p$-values).
Fisher's method still assumes the \sOneN{} per-seed $p$-values are independent, which holds because each seed is an independently seeded run, though not \emph{within} a seed's tickets.

\textbf{Bayesian supplement for the ablations.}
Each independent evolution run costs on the order of hundreds of dollars in API spend, so each ablation arm has only $n{=}5$ seeds, small enough that a $p$-value above 0.05 is easy to misread as ``no effect'' when it may just mean the test is underpowered (\S\ref{sec:threats}).
As a supplement, we instead report a Bayesian estimate of each ablation's standardized effect size, Cohen's $d$ for ablation vs.\ \setcanonical{}. We use a weakly informative prior, $\mathcal{N}(0,1)$, centered on \emph{no effect}, updated against the observed $d$ and its standard error~\cite{hedgesolkin1985} via the standard Normal-Normal conjugate rule. 
This provides a posterior over the true effect size and, from it, the posterior probability that the true effect runs in the predicted, harmful, direction. The prior pulls each posterior back toward zero relative to the raw observed effect.

\subsection{Introduction-Time Detection Dataset}
\label{sec:intro-dataset}

The \setreal{} corpus enables us to evaluate whether queries developed on clean code catch a vulnerability at the moment it is introduced. We resolve this for CVEs in our real-world corpus via SZZ-style blame attribution~\cite{sliwerski2005changes}: for each fix commit we trace its changed lines backward through \texttt{git blame} to their introducing commit, with automatic escalation from shallow to full history when a boundary is hit.

We resolve \introFiles{} files covering \introCVEs{} CVEs across \introProjects{} Python projects.
Table~\ref{tab:s5-expansion-summary} gives the per-repository breakdown. By vulnerability type across all \introProjects{} projects, we cover path traversal, XSS, code injection, SQL injection, and command injection.

For each resolved case, we export the codebase immediately before and after the introducing commit, and record which file, function, and line ranges that commit added, matching at function granularity for 109 of 133 introduced ranges and falling back to line-level for the rest. Since multiple CVEs may be introduced at different points in a repository's history, we create one snapshot per unique introducing commit.
The full per-CVE listing (repository, CWE type, commit, date, used-for-training flag) is in the released benchmark artifact.

\begin{table}[h]
\centering
\footnotesize
\begin{tabular}{@{}llcc@{}}
\toprule
\textbf{Repo} & \textbf{Type} & \textbf{CVEs} & \textbf{Caught} \\
\midrule
mistune (2 runs) & XSS/Path Trav & 4 & 4 \\
prompty & Path Trav & 1 & 1 \\
Pillow & Cmd Inj & 1 & 1 \\
langroid & Path+SQL & 1 & 1 \\
litellm (untrained) & Code Inj/Path Trav & 2 & \introCrossRepoCaught{} \\
picklescan & Code Inj & 1 & 1 \\
keras & Path Trav & 2 & 0 \\
open-webui & Path Trav & 1 & 1 \\
MobSF & XSS & 2 & 0 \\
llama\_index & SQL Inj & 1 & 0 \\
pyload & Code Inj & 1 & 1$^\dagger$ \\
changedetection.io & XSS & 2 & 0 \\
\bottomrule
\end{tabular}
\caption{Per-repository breakdown of the full \introProjects{}-project \setreal{} corpus. $^\dagger$pyload's own detector missed its CVE; it was caught only via cross-repository transfer from the open-webui and changedetection.io detectors (\S\ref{sec:real-world}). Totals: \introCVEs{} CVEs, \introCaught{} caught (\introCaughtPct{}).}
\label{tab:s5-expansion-summary}
\end{table}

\paragraph{\textbf{\setreal{} experimental protocol.}}
For each \setreal{} run, we evolve detectors on one pre-introduction tree, a repository snapshot captured immediately \textbf{before} a vulnerability-introducing commit was applied, testing whether agents can write proactive security queries on clean code that catch vulnerabilities once later introduced, without having seen any vulnerable examples.
Pre-introduction trees contain no instance of the CVE they are tested on, providing no traditional training signal. Agents must instead identify security-sensitive code patterns, such as file I/O, SQL query construction, shell command execution, path manipulation, and HTML rendering, and write queries that flag potential violations based on best practices rather than concrete examples. The fitness function for pre-introduction evolution is
$
F_{\text{proactive}} = \hat{c}_{\text{pattern}} \times (1 - \hat{r}_{\text{fp}}),
$
where $\hat{c}_{\text{pattern}}$ is the fraction of detected security-sensitive patterns covered by evolved queries, and $\hat{r}_{\text{fp}}$ the critic-judged false-positive rate of the queries' findings on the clean training corpus.

We evolve detectors on each pre-introduction snapshot, then test the resulting suite against introduction cases, mimicking a project-specific SAST tool evolving continuously and flagging newly introduced code. A detection counts as a hit if the evolved suite fires on a finding matching the reported CVE.
The resulting benchmark, \introFiles{} test cases spanning \introCVEs{} CVEs across \introProjects{} projects, is released
for evaluating any SAST tool on introduction-time recall.

\section{Results}
\label{sec:results}

We ask four questions of the evolved detectors.
The first is whether detection queries can evolve from an empty start without ground-truth labels, and they can (\S\ref{sec:trajectory}).
The second is whether the critic and its tickets behave as designed. The critic's flags are enriched for real \acp{FP} and the agent's repairs follow them above chance (\S\ref{sec:tickets}).
The third is how the queries compare with mature tools (\S\ref{sec:baselines}).
The fourth, a pilot, is whether queries written on clean code catch vulnerabilities introduced later (\S\ref{sec:real-world}).

\subsection{RQ1: Can static-analysis queries evolve without ground-truth labels, from scratch?}
\label{sec:trajectory}

\begin{figure*}[htbp!]
    \centering
    \includegraphics[width=0.90\linewidth]{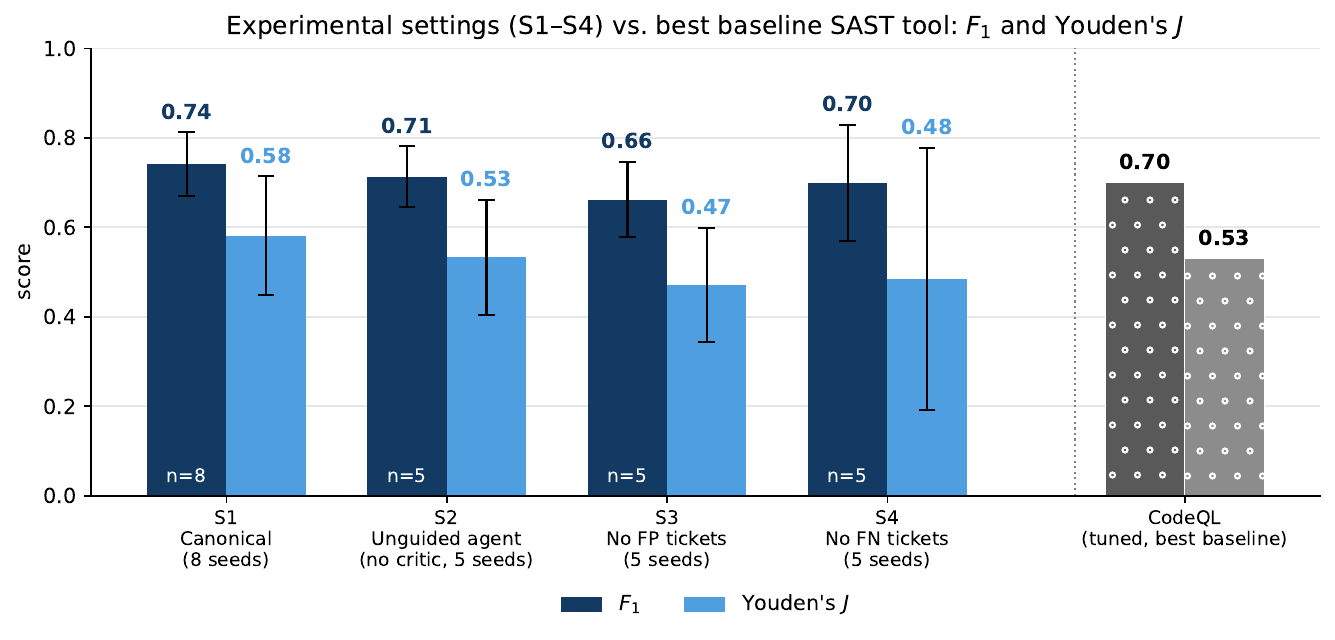}
    \caption{\textbf{Ground-truth performance by setting.}
    \setcanonical{} matches the best baseline tool, and removing any one mechanism lowers the mean on both metrics, though not significantly at this sample size (\S\ref{sec:trajectory}).
    Ground-truth \fscore{} and Youden's $J$ by setting, against the best baseline tool, mean $\pm$ 95\% CI over independent seeds.
    Youden's $J=$ sensitivity $+$ specificity $-1$; for the evolved detector, \acp{TN} are approximated as corpus-safe files minus \acp{FP} (an upper bound on $J$), whereas the baseline tool's $J$ is computed from exact \acp{TN}, so the two are not on strictly equal footing.}
    \label{fig:combined-results}
\end{figure*}

We run \sOneN{} independent evolution sessions from an empty starting point, varying only the random seed, and score each returned suite against the withheld labels.
Precision is the share of a suite's alerts that are real, and recall is the share of the benchmark's vulnerabilities it finds.
The mean \fscore{}, which balances the two, is $\sOneFmean{}$, from a mean precision of $\sOnePrecMean{}$ and a mean recall of $\sOneRecMean{}$.
The runs differ because the loop is stochastic and we do not control decoding settings, with a standard deviation of $\sOneFsd{}$ and a 95\% confidence interval of $[\sOneFciLo{}, \sOneFciHi{}]$.
Against the strongest baseline, \sOneBeatBaseline{} of \sOneN{} seeds exceed CodeQL's \fscore{} of \codeqlf{} (Figure~\ref{fig:combined-results}), and the other two fall below it, so the mean sits only modestly above the baseline and its interval overlaps CodeQL's score.

Figure~\ref{fig:combined-results} shows the three ablations against \setcanonical{}.

\paragraph{\textbf{\setunguided{}: removing the critic raises the false-positive rate, without moving \fscore{} significantly.}}
With no critic output, precision falls from $\sOnePrecMean{}$ to $\sTwoPrecMean{}$ ($p=\sOneVsTwoPrecp{}$) as the agent writes looser detectors with nothing to prune \acp{FP}, while recall rises from $\sOneRecMean{}$ to $\sTwoRecMean{}$ ($p=\sOneVsTwoRecp{}$).
The \fscore{} drop does not reach significance ($p=\sOneVsTwoFp{}$) at this sample size because the precision loss and recall gain partly cancel in the harmonic mean.

\paragraph{\textbf{\setnofp{}: removing FP tickets trends worse on every metric.}}
\setnofp{} is the closest of the three ablations to significance on \fscore{} ($p=\sThreeVsOneFp{}$).
Youden's $J$ ($p=\sThreeVsOneJp{}$), precision ($p=\sThreeVsOnePrecp{}$), and recall ($p=\sThreeVsOneRecp{}$) all trend the same direction, consistent with the \ac{FP}-ticket mechanism costing detector quality broadly rather than trading one axis for another.

\paragraph{\textbf{\setnofn{}: removing FN tickets leaves the mean unchanged but widens the spread.}}
Only the pointer to likely-missed vulnerabilities is removed; the critic's precision estimate, \ac{FP} tickets, and fitness score are unchanged.
Precision and recall drift the same way as in \setunguided{} but more weakly (recall $p=\sOneVsFourRecp{}$, precision $p=\sOneVsFourPrecp{}$), and the \fscore{} difference is not significant ($p=\sOneVsFourFp{}$).

None of the three individually establishes that the mechanism drives \fscore{}, so we treat them as directionally consistent, not as proof.
Pooling all three via Fisher's combined probability test (\S\ref{sec:tickets}'s pooling method) leaves that conclusion unchanged on \fscore{} ($\chi^2=\ablFisherFstat{}$, $\mathit{df}=6$, $p=\ablFisherFp{}$) but moves closer to significance on precision ($\chi^2=\ablFisherPrecStat{}$, $p=\ablFisherPrecp{}$), a real but modest effect that three $n{=}5$ ablations remain underpowered to establish alone.

As \S\ref{sec:statistical-tests} describes, we supplement the frequentist tests with a Bayesian effect-size estimate under a skeptical prior. 
\setnofp{} keeps most of its signal: from $d=\ablBayesThreeD{}$ to $d=\ablBayesThreePostD{} \pm \ablBayesThreePostSd{}$, with \ablBayesThreeProb{} posterior probability the effect is real and harmful. 
\setunguided{} and \setnofn{} shrink further, landing at more modest posterior effects ($\ablBayesTwoPostD{} \pm \ablBayesTwoPostSd{}$, \ablBayesTwoProb{}; $\ablBayesFourPostD{} \pm \ablBayesFourPostSd{}$, \ablBayesFourProb{}). 
The controlled evidence that the critic/ticket mechanism does what it is designed to do comes from directly testing its behavior, not from these outcome ablations. \S\ref{sec:tickets} does that next.

\begin{pilotbox}
\textbf{RQ1 takeaway.} Evolution from a stripped, empty workspace with no ground-truth signal at any point reaches mean \fscore{} $\sOneFmean{} \pm \sOneFsd{}$ across \sOneN{} independent runs, above CodeQL's tuned, \emph{label-informed} \fscore{} of $\codeqlf{}$ on \sOneBeatBaseline{} of \sOneN{} seeds.
The no-critic ablation reaches $\sTwoFmean{}$, with higher recall ($\sTwoRecMean{}$) and lower precision ($\sTwoPrecMean{}$) than \setcanonical{}.
The three ablations trend in the predicted direction and none is significant on \fscore{}.
\end{pilotbox}

\subsection{RQ2: Does the ticket mechanism behave as designed?}
\label{sec:tickets}

The critic opens a ticket for each finding it doubts, and the agent either repairs the query so the finding disappears or contests the ticket (\S\ref{sec:system}).
We check this mechanism against the withheld ground truth across all \tkMultiN{} \setcanonical{} seeds.
We ask whether the critic flags findings that really are false alarms, whether the agent repairs those more often than chance would predict, and whether the critic's estimate of its own precision is fair.
Exact tests answer each in turn (\S\ref{sec:ticket-method}, \S\ref{sec:statistical-tests}).

\textbf{The critic's tickets are enriched for real \acp{FP}, and the agent acts on that signal.} Removing a ticketed finding that is a \ac{FP} raises \fscore{}, but removing one that is a real vulnerability lowers it, so we focus here on whether the critic identifies genuine \acp{FP}, and whether the agent repairs those more than it would by chance.

\paragraph{A concrete example.}
The critic opened a ticket for \texttt{random.randbytes(32)} generating a \texttt{remember\allowbreak Me} session cookie, judging 256 bits of PRNG output sufficient with no attacker-controlled seed apparent. The agent contested. Absent an explicit \texttt{SystemRandom} instance, \texttt{random.randbytes()} relies on Python's Mersenne Twister, the exact CWE-330 pattern the benchmark's sibling \texttt{weakrand} cases target. The critic re-judged, agreed, and reversed the ticket to confirmed vulnerability (Figure~\ref{fig:ticket-dialogue}). 
This exchange is taken from one archived run. 
The same pattern of doubt, repair or contest, and resolution recurs across the \tkMultiTickets{} tickets summarized below, most resolving with less back-and-forth.

\begin{widefig}[hbt!]
\centering
\resizebox{0.9\linewidth}{!}{%
\begin{tikzpicture}[
  font=\tiny,
  bubble/.style={rectangle, rounded corners, draw, text width=42mm, align=left, inner sep=1.3mm},
  critic/.style={bubble, fill=nbnavy!10, draw=nbnavy},
  agent/.style={bubble, fill=nblightblue!15, draw=nblightblue!80!black},
  arr/.style={-{Stealth[length=1.4mm]}, thick, gray},
]
\node[critic] (c1) at (0,0)
  {\textbf{Critic} (initial)\\ \textit{``256-bit PRNG output; no attacker-controlled seed visible --- insufficient evidence.''}};
\node[agent, right=6mm of c1] (a1)
  {\textbf{Agent} (contest)\\ \textit{``\texttt{randbytes()} draws from the Mersenne Twister unless \texttt{SystemRandom} is used --- CWE-330, the exact \texttt{weakrand} pattern.''}};
\node[critic, right=6mm of a1] (c2)
  {\textbf{Critic} (re-judged)\\ \textit{``Agreed: MT19937 output is predictable; the cookie is a security-sensitive sink. Verdict flips \acs{FP}$\to$real.''}};
\draw[arr] (c1) -- (a1);
\draw[arr] (a1) -- (c2);
\end{tikzpicture}%
}
\caption{\textbf{The exchange behind the ticket in the concrete example above}: the critic's first read, the agent's contest, and the critic's reversal.}
\label{fig:ticket-dialogue}
\end{widefig}

Across all \tkMultiN{} \setcanonical{} seeds, the agent opened \tkMultiTickets{} tickets covering \tkMultiFindings{} distinct findings and repaired \tkMultiSolved{} of them, where a ticket counts as solved once the agent's repair removes its finding.
Of those, \tkMultiSolvedFP{} were genuine \acp{FP} and \tkMultiSolvedReal{} were real vulnerabilities removed by mistake, a pooled solve precision of \tkMultiPrecisionPooled{}.
Seed by seed, the precision is $\tkMultiPrecisionMean{} \pm \tkMultiPrecisionSd{}$, with a range of \tkMultiPrecisionMin{}--\tkMultiPrecisionMax{}.

\begin{widefig}[hbt!]
\includegraphics[width=\widefigw]{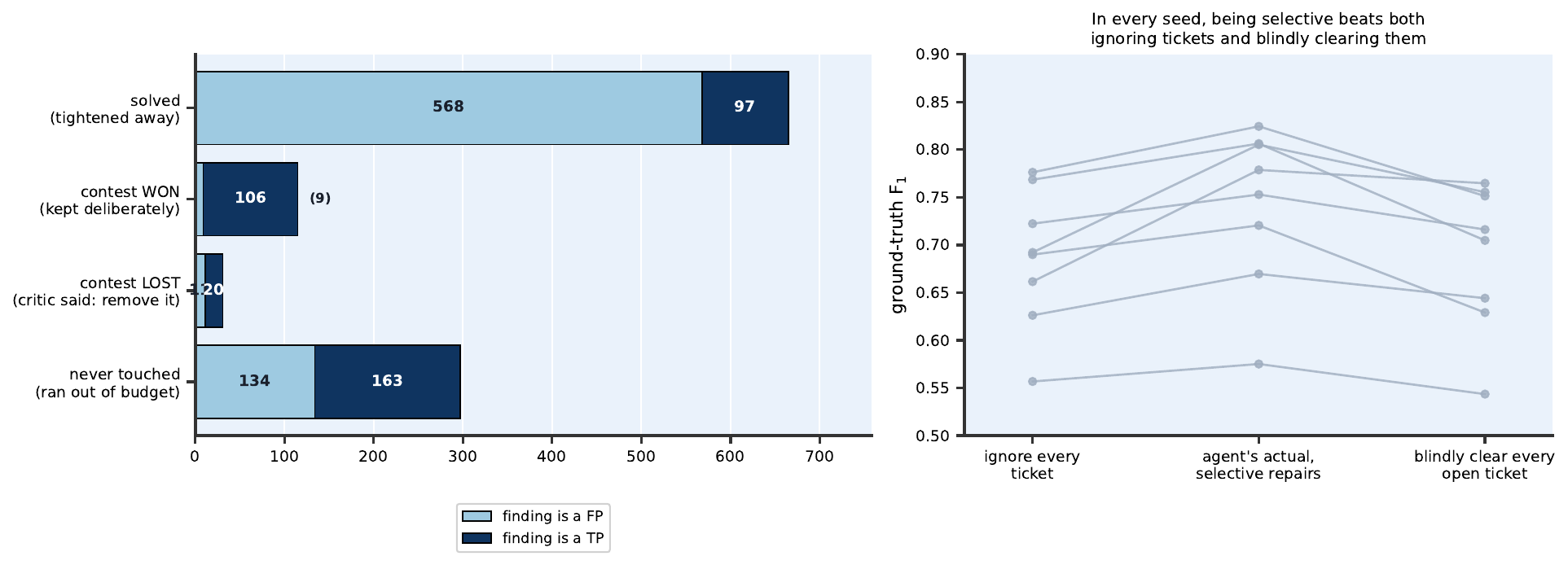}
\caption{Ticket mechanism outcomes, pooled and per-seed across all \tkMultiN{} \setcanonical{} runs, scored against the withheld labels.
\textbf{Left:} outcome for each ticketed finding by ground-truth label; of \tkMultiSolved{} findings solved in total, \tkMultiSolvedFP{} were genuine \acp{FP} and \tkMultiSolvedReal{} were real vulnerabilities removed by mistake.
\textbf{Right:} ground-truth \fscore{} under three repair strategies (ignore every ticket, the agent's actual selective repairs, blindly clear every open ticket), plotted per seed; in every seed, being selective beats both extremes.}
\label{fig:ticket-decisions}
\end{widefig}

Across all \tkMultiN{} runs, blindly clearing every open ticket (fixing \acp{FP} and real vulnerabilities alike, with no selectivity) scores \emph{below} the agent's actual, selective repairs (Figure~\ref{fig:ticket-decisions}).

\textbf{Targeting.} A ticketed finding is more likely to be a genuine \ac{FP}. 
A per-seed Fisher's exact test is significant on \tkMultiTargetSig{}/\tkMultiN{} seeds, and combining all \sOneN{} ($p=\tkMultiTargetP{}$) rejects the null that tickets are independent of \ac{FP} status.

\textbf{Selection.} The agent's repair choice tracks that signal rather than ignoring it. 
A per-seed one-sided hypergeometric test is significant on \tkMultiSelectSig{}/\tkMultiN{} seeds, and combined ($p=\tkMultiSelectP{}$) shows the agent repairs more genuine \acp{FP} than chance would predict, listening to the critic in every run.

\textbf{Calibration.} The critic's own precision estimate is conservative.
Across all \tkMultiCalibPos{} seeds, the direction never flips, underrating its detectors on average (\tkMultiCalibGapMean{}, SD \tkMultiCalibGapSd{}, range \tkMultiCalibGapMin{}--\tkMultiCalibGapMax{}), though only \tkMultiCalibSig{}/\tkMultiN{} of the per-seed Wilcoxon tests clear conventional significance alone.

Across \sOneN{} independent trajectories, the critic's tickets point disproportionately at genuine \acp{FP}, the agent acts on that signal, and where the critic is wrong about its own accuracy, it errs on the safe side.

\textbf{False-negative tickets.} The \ac{FN} tickets do the same job for recall.
A one-sided Fisher's exact test is significant on \tkMultiFNSig{}/\tkMultiN{} seeds, pooling to \tkMultiFNn{} FN tickets, of which a pooled share of \tkMultiFNprec{} point at a genuine miss (mean \tkMultiFNPrecMean{} $\pm$ \tkMultiFNPrecSd{} averaged seed by seed), pointing the agent toward the higher-risk part of the corpus despite seed-to-seed variance.

\begin{pilotbox}
\textbf{RQ2 takeaway.} Ticketed findings are enriched for genuine \acp{FP} (Fisher's exact, $p=\tkMultiTargetP{}$), and the agent's repairs track that signal above chance ($p=\tkMultiSelectP{}$), across all \sOneN{} seeds.
Of the \tkMultiSolved{} solved findings, \tkMultiSolvedReal{} were real vulnerabilities removed by mistake.
Where the critic misjudges its own precision, it underrates rather than overrates.
\end{pilotbox}

\subsection{Agent Behaviour}
\label{sec:behaviour}

Across the $\hToolCalls{}$ tool calls of the committed generation, the agent edits existing detectors ($\hEditQuery{}$ calls) far more often than it writes new ones from scratch ($\hWriteQuery{}$), and spends the bulk of its budget reading and exploring the corpus before touching a query at all. Figure~\ref{fig:gen-behaviour} shows this is not one habit sustained throughout but a rhythm repeated every generation: explore and read, then a burst of editing, then an evaluation call that closes the loop and starts the next round.

\textbf{Compute cost.}
One \setcanonical{} run uses around $\costOneAgentIn{}$ input and $\costOneAgentOut{}$ output \sonnet{} tokens for the agent, plus at least $\costOneCriticIn{}$ input and $\costOneCriticOut{}$ output \opus{} tokens for the critic.

\begin{widefig}[hbt!]
\includegraphics[width=\widefigw]{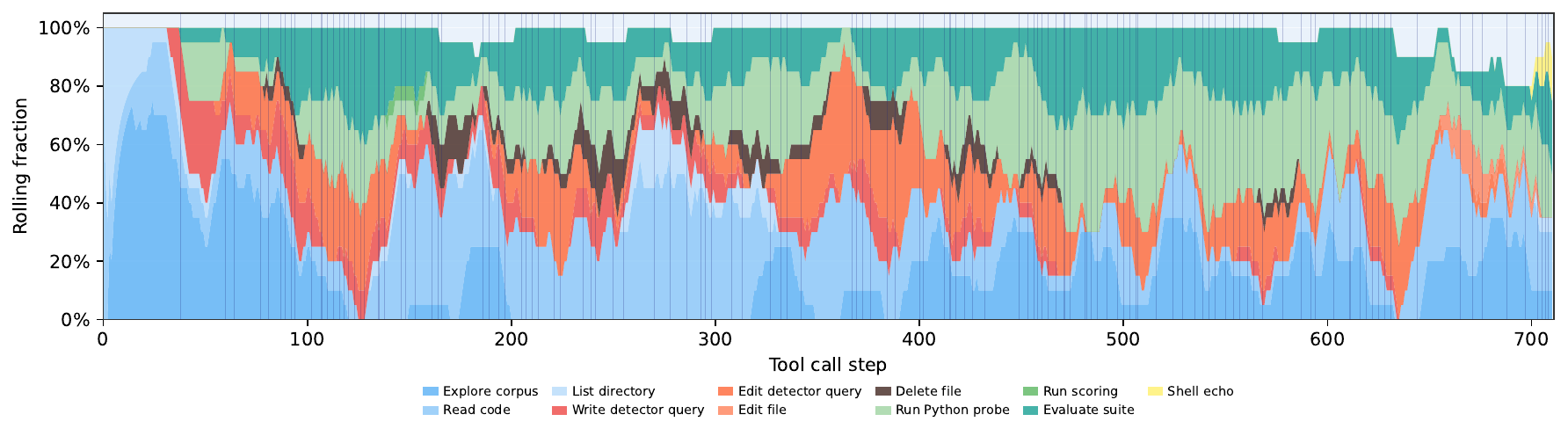}
\caption{Agent activity composition across the committed generation (single session; commands classified by shape). Faint vertical ticks mark each evaluation against the corpus.}
\label{fig:gen-behaviour}
\end{widefig}

\subsection{RQ3: Do static-analysis queries evolved without ground-truth labels reach the quality of mature, hand-maintained SAST tools?}
\label{sec:baselines}

A detector that only beats an empty baseline is not useful in practice, so we compare against the tools practitioners already use.
The panel is five widely used tools (CodeQL, Semgrep, Bandit, Bearer, OWASP ZAP), each tuned with access to the withheld labels that \sysname{} never sees, plus a one-shot LLM-written \joern{} baseline that shows what the evolution loop adds (\S\ref{sec:config}).
Each tool contributes a single best-of-9 score, and its tuning budget is not matched to \sysname{}'s.

\begin{widefig}[hb!]
\centering

\begin{subfigure}[t]{0.49\linewidth}
    \centering
    \includegraphics[width=\linewidth]{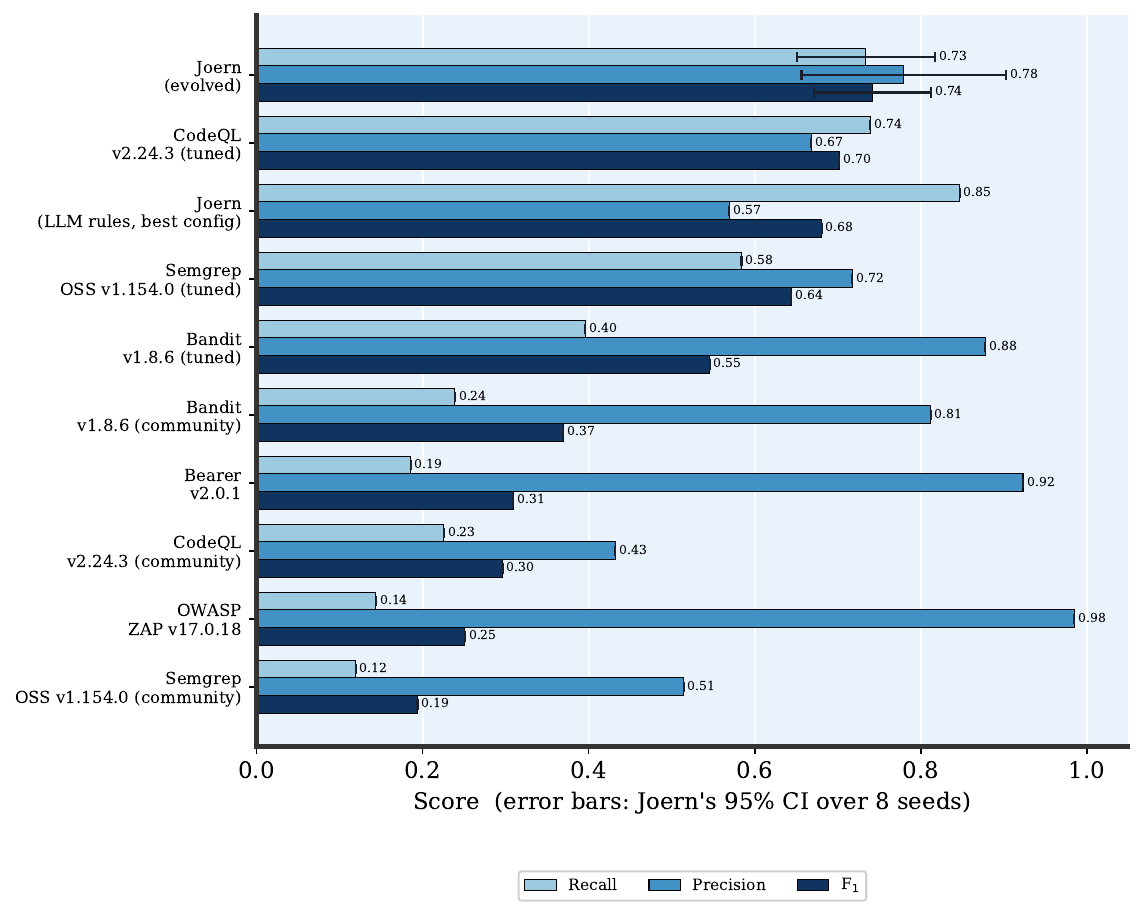}
    \caption{Performance breakdown, tuned and community configurations.}
    \label{fig:baseline-breakdown}
\end{subfigure}
\hfill
\begin{subfigure}[t]{0.49\linewidth}
    \centering
    \includegraphics[width=\linewidth]{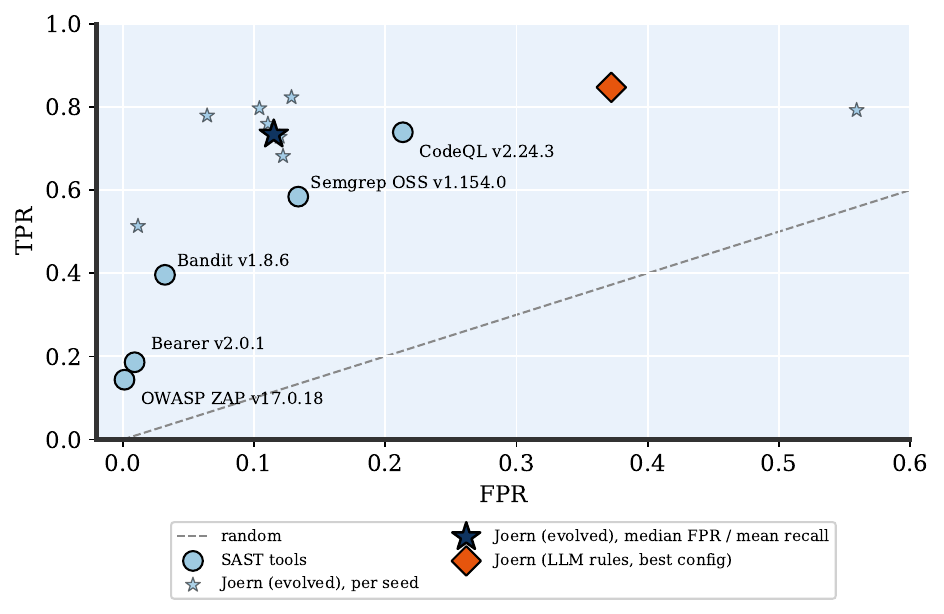}
    \caption{Recall versus \ac{FPR}.}
    \label{fig:baseline-recall-fpr}
\end{subfigure}

\caption{\textbf{Comparison with baseline tools.}
The evolved detector's precision-recall tradeoff sits closest to the upper-left of any tool in the panel, driven mainly by precision rather than recall (\S\ref{sec:baselines}).
(a) Recall, precision, and \fscore{} for each tool; CodeQL, Semgrep, and Bandit show both community and tuned configurations (Bearer/ZAP ship one configuration; \joern{} has no out-of-the-box ruleset, so only its best one-shot LLM configuration is shown).
(b) Recall versus \ac{FPR} for each tool's strongest (tuned) configuration; the evolved \joern{} detector is plotted once per \setcanonical{} seed (light stars) and once as a median-\ac{FPR}/mean-recall summary (dark star); the diamond marker is the best one-shot LLM-written \joern{} ruleset. Points closer to the upper-left are preferable.
}
\label{fig:baseline-comparison}
\end{widefig}

The evolved detector reaches the highest mean \fscore{} of the panel ($\sOneFmean{} \pm \sOneFsd{}$), ahead of tuned CodeQL ($\codeqlf{}$), the strongest baseline, and of every other tool (Figure~\ref{fig:baseline-comparison}), including the one-shot \joern{} ruleset ($\joernllmf{}$).
The margin over CodeQL is small and the \sOneN{}-seed 95\% confidence interval overlaps CodeQL's score, so the fair reading is that the evolved queries match the best tuned tool without ever seeing a label.

The advantage over CodeQL comes mainly from precision.
Mean recall ($\sOneRecMean{}$) is close to CodeQL's ($\codeqlrecall{}$), while mean precision is $\sOnePrecMean{}$, and the two highest-precision tools, Bearer and ZAP, reach theirs by giving up most of their coverage.
The community, out-of-the-box configurations of CodeQL, Semgrep, and Bandit score well below their tuned versions ($\codeqlfstock{}$, $\semgrepfstock{}$, $\banditfstock{}$), and the evolved detector clears them by a wide margin.

\begin{pilotbox}
\textbf{RQ3 takeaway.} Tuned with access to the labels, CodeQL scores $\codeqlf{}$, and the evolved detector, which never sees a label, scores $\sOneFmean{}$ on average, a difference within the seed-to-seed interval.
It is ahead of every other tuned tool by a wider margin and of every community configuration by a wide one.
\end{pilotbox}

\subsection{RQ4 (pilot study): Can agents write proactive security rules on clean code that catch vulnerabilities at introduction time?}
\label{sec:real-world}

Previous RQs have focused on reactive detection.
We now ask whether an agent in the same setting can write detectors that catch vulnerabilities proactively, at the moment they are introduced.
We ask this as a question about the agent's zero-shot reasoning, not about whether \sysname{}'s evolutionary fitness mechanism specifically drives it: the two come apart in this pilot, and we treat that separation as a finding in its own right. 
 
We evolve detectors on real-world repositories' pre-introduction snapshots, using the pattern-coverage fitness of \S\ref{sec:intro-dataset} (no precision/recall signal is available on a corpus with zero vulnerabilities by construction), then test the resulting suite against the actual introducing commit.

We report this as a pilot study on a benchmark we release for it. 
Zero-shot agentic reasoning caught \introCaught{} of the \introCVEs{} CVEs in this sample. 

Each detector is tested only on introduction cases from the repository whose snapshot it was evolved on, except where noted as cross-repository transfer.

\paragraph{\textbf{The capability result: detection succeeded more often than not.}}
Across all \introRuns{} runs, evolved detectors caught \introCaught{} of \introCVEs{} CVEs.
Table~\ref{tab:s5-expansion-summary} gives the full per-repository breakdown. \introInRepoCaught{} of the \introCaught{} catches came from a detector evolved on the same repository, out of \introInRepoTotal{} CVEs in trained repositories. litellm, the one repository without a training run, was missed on both of its CVEs, and keras, MobSF, llama\_index, and changedetection.io caught none of their own CVEs. 
The remaining catch is a cross-repository transfer: pyload's own detector missed its CVE while the open-webui and changedetection.io detectors both flagged it, direct evidence that a rule evolved on one codebase can generalize to a vulnerability pattern in an entirely different one.

The lowest-scoring repositories are also among the largest and most complex (llama\_index has 3{,}893 files, while keras and MobSF each run in the hundreds), while picklescan, the smallest at 7 files, was detected perfectly. We flag this only as an observed correlation across a 12-repository sample, not a tested hypothesis, since we have not measured what drives it.


\begin{pilotbox}
\textbf{RQ4 takeaway.}
Zero-shot agentic reasoning caught \introCaught{} of \introCVEs{} introduction-time CVEs across \introProjects{} real-world repositories (95\% CI \introCaughtCI{}), including one catch by detectors evolved on other repositories. Separately, the fitness signal we designed to drive this setting was inert for most runs, and the catch rate varied heavily by repository.
\end{pilotbox}

\section{Related Work}
\label{sec:background}

\sysname{} sits between several ways of finding software vulnerabilities, and inherits from each without fully belonging to any of them.
Static analyzers such as \joern{}~\cite{yamaguchi2014, cpg2014}, CodeQL~\cite{codeql}, Semgrep~\cite{semgrep}, and Infer~\cite{infer} expose programmable queries over syntax, control flow, and data flow, and stay deterministic and explainable only as long as those queries keep up with how languages and frameworks actually change, a maintenance burden that grows with deployment scale~\cite{bessey2010few}.
\sysname{} keeps that same deterministic query as its output, but rewrites it continuously, guided by a critic's pseudo-labels rather than a human maintainer.

A parallel line of work sidesteps rule maintenance by learning to classify vulnerable code directly: Devign~\cite{devign}, SySeVR~\cite{sysevr}, ReVeal~\cite{reveal}, LineVul~\cite{linevul}, and pretrained encoders such as GraphCodeBERT~\cite{graphcodebert} all reach strong predictive accuracy this way, but the result is a classifier rather than an analysis, and what these models learn stays in their weights rather than in a query a security engineer can read, edit, or run on its own.

\textbf{LLM agents for software security.}
LLM agents push further in the same direction, applying language models directly to vulnerability detection~\cite{gao2023farhavewegone,ding2024howfararewe,sun2024llm4vuln, absentia}, code review~\cite{cihan2025codereview}, repair~\cite{kulsum2024vrpilot}, exploit generation~\cite{fang2024exploit}, and penetration testing~\cite{deng2023pentestgpt}, with agentic tool use adding iterative execution feedback~\cite{yao2022react,yang2024sweagent,autocoderover}; Codebadger, for instance, exposes \joern{}'s \ac{CPG} to an agent through slicing and taint-tracking tools rather than raw queries~\cite{10.1145/3786165.3788441}.
Here the model is doing the analysis itself, on every repository and every run, so little of what it learns carries over to the next one, even where a second model checks its work, as in recent verifier-style vulnerability discovery~\cite{vecho2026}.
\sysname{} also uses a second model this way, as an \ac{LLM}-as-judge~\cite{zheng2023judge}, but points the resulting signal at the query rather than the verdict, so what persists across runs is the analyzer rather than a transcript.

\textbf{LLM-generated analysis rules.}
A closer line of work has \acp{LLM} write the rules themselves.
QLCoder synthesizes CodeQL queries from CVE metadata and vulnerable and patched code, refining each query with execution feedback~\cite{qlcoder2025}.
Irsan et al. generate CodeQL queries from National Vulnerability Database entries~\cite{irsan2026querygen}, and MoCQ refines LLM-written vulnerability patterns against symbolic validation feedback~\cite{mocq2025}.
These systems take their supervision from outside, as CVE reports, patches, or pattern specifications, whereas \sysname{} takes it from a critic's pseudo-labels over unlabeled code and keeps evolving the same detector across generations.

\textbf{Self-improving software engineering agents.}
Reflexion~\cite{reflexion}, Self-Refine~\cite{selfrefine}, TextGrad~\cite{textgrad}, OPRO~\cite{opro}, and GEPA~\cite{gepa} face a related problem from the agent's side: they get better at reasoning through refined prompts, strategies, or policies, but that improvement stays embedded in the model pipeline and inherits continual learning's stability-plasticity dilemma~\cite{grossberg1980does,mermillod2013stability}.

\textbf{Evolutionary search and neuroevolution.}
The archive-and-mutation loop underneath \sysname{} is closer to genetic programming's evolution of executable programs~\cite{koza1992genetic} and neuroevolution's search beyond fixed parameters~\cite{galvan2020neuroevolution,leclei2023neuroevolution}, both of which justify branching, non-greedy search over an irregular landscape.
Recent surveys extend that idea to \ac{LLM}-based systems~\cite{selfevolvingagents2026}: Meta-Harness evolves the harness around a fixed model~\cite{metaharness2026}, and HarnessX treats the harness itself, prompts, tools, memory, control flow, as a composable artefact~\cite{harnessx2026}.
\sysname{} shares the premise that the machinery around a model is worth evolving, but points it at the analysis logic itself, selected by a critic-pseudo-labeled signal that needs no ground truth, with a feedback step that draws on Reflexion, Self-Refine~\cite{reflexion,selfrefine}, and GEPA's critique/generation split~\cite{gepa}.

\section{Discussion}
\label{sec:discussion}


The critic is the agent's only source of feedback.
The tickets are enriched for genuine \acp{FP}, and the agent repairs those more often than chance would predict.
Both effects are significant across seeds.
The outcome ablations point the same way. 
Removing the critic, the \ac{FP} tickets, or the \ac{FN} tickets each lowers the mean \fscore{}, and the Bayesian estimate puts \ablBayesThreeProb{} posterior probability on the \ac{FP}-ticket removal being harmful.
These differences are not individually significant, so we read them as a consistent direction whose size a larger campaign would pin down.
At the same time, the agent writes strong detectors without the critic.
The critic's contribution is therefore a refinement on top of a capable agent, and removing it trades precision for recall.

The ticket tests should be read the same way.
Ticketed findings are more likely than other findings to be genuine \acp{FP}, and the agent repairs those more often than chance, but they establish enrichment and not accuracy: \tkMultiSolvedReal{} of the \tkMultiSolved{} findings the agent solved were real vulnerabilities removed by mistake.
The \ac{FN} tickets show the same limit, since a pooled \tkMultiFNprec{} of them point at a genuine miss.
The contest step lets the agent push back on a wrong verdict, as in the \texttt{randbytes} example of \S\ref{sec:tickets}, but it relies on a single critic.
A natural next step is to separate the judgments, for example with independent critics for \ac{FP} and \ac{FN} review that must agree before a ticket is opened, to make this feedback more reliable.


For a security team, the practical result is that the maintenance loop can run without labels.
A team with a new framework or a codebase no one has labeled could point the agent at it and review the queries it returns, instead of writing them.
We tested this on a labeled benchmark and a small pilot, so the claim is that the loop can work without labels, not that it has been shown to work on unlabeled production code.

Two properties of the output matter for that use.
First, the cost is paid once and not on every scan.
A run consumes substantial model tokens (\S\ref{sec:behaviour}), but the suite it returns is ordinary code with no model dependency, so scanning afterwards is as cheap as running any \joern{} query set, unlike approaches that call a model on every repository and every run.
Second, the suite can be reviewed.
The agent often records why a flagged case is safe in an inline comment (\S\ref{sec:example}), so an engineer can audit a change, keep it, or revert it, which keeps a person in the loop without asking them to write the rule.


The agent's only memory across generations is a flat, read-only archive of earlier suites.
Structured memory~\cite{wang2024agentworkflowmemory,chhikara2025mem0} could let it reuse what it has learned instead of re-deriving it after each repair.
Replacing the stronger critic with the agent's own model would also show how much of the signal comes from the critic's prior knowledge.

\section{Threats to Validity}
\label{sec:threats}

We organize the threats along four axes, and for each we point to the evidence in \S\ref{sec:results} that bounds it.

\textbf{Internal validity.}
The agent explores the corpus directly, so some of its gains may come from its own reading of the code and not from the critic's signal.
The \setunguided{} ablation bounds this, since it keeps the agent's tools and corpus and removes only the critic, and it lands within \setcanonical{}'s spread on \fscore{}, so the outcome data cannot separate the two.
The critic is also a stronger model than the agent it supervises (\S\ref{sec:models}), so some of what we attribute to pseudo-labeling could be the critic's own prior knowledge leaking through its judgments.
Isolating that requires a same-model agent and critic, a local-context-only critic, and a category-blind critic, which we have not run at the cost \S\ref{sec:behaviour} reports.
For the real-world pilot, the ground truth is the introducing commit found by SZZ attribution, which attributes lines to their last modification and can misplace the true introduction.

\textbf{External validity.}
We evaluate one language (Python) and one model pair, and the OWASP benchmark is synthetic, so we do not claim the results carry to other languages, model families, or real-world development.
The real-world pilot offers some evidence for transfer to organically written code, but it covers \introProjects{} projects and \introCVEs{} CVEs.
\joern{}'s \ac{CPG} construction is bound by available RAM, so scaling to large repositories or other program representations is open, and a single \setcanonical{} run already consumes a substantial number of \sonnet{} tokens (\S\ref{sec:behaviour}).
We did not study other budgets, so we cannot say how detector quality or cost scale with a smaller or larger one.

\textbf{Construct validity.}
The pseudo-labeled fitness is a proxy for the ground-truth \fscore{} we cannot observe during search, and its two halves are each imperfect.
``Label-free'' means the agent never sees which files are vulnerable, not that the corpus offers no prior structure.
The target taxonomy $T$ is fixed (for OWASP, the benchmark's own \ac{CWE} categories), and \joern{}'s \ac{CPG} exposes taint-relevant primitives that narrow what a plausible query can look like.
Both models were also pretrained on public code and vulnerability discussion that may include the OWASP style, and we did not test the critic under semantic-preserving transformations that remove those cues.
Some reported accuracy may therefore reflect benchmark recognition and not general detection ability.
Agent and critic also come from one model family, which may make the critic favor output that resembles its own~\cite{panickssery2024selfpref}.
Youden's $J$ for the evolved detector uses an approximation of \acp{TN} that upper-bounds it, so it is not on equal footing with the baselines' $J$.
The evidence therefore supports a narrower claim than ``without labels.''
The agent has no access to the withheld answer key, on a benchmark whose taxonomy and surface patterns are not entirely unfamiliar to the search.

\textbf{Conclusion validity.}
The loop is stochastic, and ground-truth \fscore{} varies across independent runs (\S\ref{sec:trajectory}), with the charted run sitting above the mean.
Each ablation has five seeds, which bounds the aggregate metric and does not show the mechanisms are inert, and a conventionally powered campaign would need roughly ten times as many, which is why we add a Bayesian effect-size estimate.
The ablation tests are not corrected for multiplicity, and \fscore{} is the single primary metric.
The ticket tests are significant across all \sOneN{} seeds, but they come from one benchmark and one model pair, so they establish that the critic tracks ground truth in this setting and not in general.

\section{Conclusion}
\label{sec:conclusion}

We asked whether a static-analysis rule set can evolve and repair itself without ground-truth labels, the way a security engineer revises a set of queries after spotting what it missed.
Our results support three findings, with the limits stated below.
First, an agent can write competitive queries from scratch.
They reach a mean \fscore{} of $\sOneFmean{}$ on the OWASP Benchmark for Python, ahead of five widely used tools' default configurations and in the same range as CodeQL once tuned with the labels \sysname{} never gets.
Second, the critic's pseudo-labels carry a real but imperfect signal.
Its tickets are enriched for \acp{FP} and the agent's repairs track them above chance, although some repairs remove real vulnerabilities.
Removing the critic or either ticket type lowers the mean \fscore{} in the direction the design predicts.
Third, an agent reasoning over clean code can write rules that catch later vulnerabilities.
Queries evolved on clean code alone caught \introCaught{} of \introCVEs{} vulnerabilities at the moment they were introduced.
We hope these findings help reframe the security scanner from a static tool into a system that keeps itself current as the code it watches keeps changing.
\section*{Data Availability}
We provide an anonymized repository (\url{https://anonymous.4open.science/r/SESAA}).

\section*{Generative AI Disclosure}
Per ACM's policy on authorship, we disclose the use of generative AI tools in this work. Claude Code assisted with implementing and modifying portions of the \sysname{} system and analysis scripts, and with drafting and revising portions of this manuscript. ChatGPT assisted with drafting and editing portions of the manuscript text. Perplexity was used for literature search, background research and iterative manuscript revision.

\section*{Aknowledgments}
This work was conducted as part of the Amazon Nova AI Challenge, which provided funding and AWS compute credits to participating university teams. The study design, model selection, analysis, and the decision to publish were made independently by the authors.
\clearpage
\bibliographystyle{plainurl}
\bibliography{references}
\clearpage
\clearpage
\pagebreak
\appendix

\section{Agent Tools}
\label{sec:appendix-tools}

Each generational agent is a \acReAct{} loop equipped with the six tools below (seven when the \ac{FP}-ticket mechanism is enabled, \S\ref{sec:fitness}) and no other side-effecting capabilities. In particular, \texttt{bash} executes inside the sandboxed Docker analysis container, where \texttt{/repo} is the agent's mutable workspace and \texttt{/corpus} is the read-only OWASP corpus, and is blocked from reading the withheld ground-truth labels or the critic implementation.

\begin{description}
\item[\texttt{think(reasoning)}] A reasoning checkpoint with no side effects. The agent records its plan (which false negatives to target, what mutation to attempt) before acting; this keeps the trajectory focused but does not alter any state.

\item[\texttt{bash(command)}] Runs a shell command in the analysis container. Used to explore the corpus, \texttt{grep} for patterns, and write detector files (\texttt{.sc} queries or \texttt{.py} tools) via here-documents. Commands that reference the ground-truth directory or critic files are rejected, as is \texttt{sleep}.

\item[\texttt{analyse\_corpus()}] Compiles and runs all custom detectors over the full corpus ($\sim$2~min) and returns a per-query findings summary together with the critic's qualitative feedback (coverage gaps, false-positive patterns). It measures the current suite but does not commit it or end the generation. The agent may call it repeatedly, revising detectors between calls, before deciding the suite is ready (\S\ref{sec:config}).

\item[\texttt{submit()}] Commits the current suite as this generation's candidate and ends the agent's turn, so each generation yields exactly one committed candidate (\S\ref{sec:results}) regardless of how many times \texttt{analyse\_corpus()} was called beforehand.

\item[\texttt{delete\_query(name)}] Removes a detector file the agent previously wrote. Because a single non-compiling query makes \texttt{analyse\_corpus()} fail for \emph{all} queries, deleting an unfixable one unblocks the rest.

\item[\texttt{inspect\_prior\_candidates(candidate\_id)}] Read-only access to the archive of past candidates and their critic evaluations (reasoning, coverage gaps, false-positive patterns), giving the agent non-Markovian memory across generations. Called with no argument it lists all candidates; with an ID it returns that candidate's full critic evaluation.
\end{description}

\section{Prompts}
\label{sec:appendix-prompts}

We reproduce verbatim the four prompts that define the system's behaviour: the three agent-facing prompts (system, per-generation, and the critic feedback injected back into the agent), plus the critic's precision-judge prompt that computes the precision half of the fitness signal.
Template variables (e.g.\ \texttt{\{gen\}}, \texttt{\{parent\_id\}}) are filled at runtime.

\subsection{System Prompt}
\label{sec:prompt-system}

The system prompt establishes the agent's role, workspace layout, available tools, and the measure/refine/commit loop; \S\ref{sec:config} describes the same mechanism in prose.

\begin{lstlisting}[style=python,language={},basicstyle=\ttfamily\scriptsize,caption={System prompt (verbatim; the \texttt{ticket(...)} tool is included only when the \ac{FP}-ticket mechanism is enabled, \S\ref{sec:fitness}).},label={lst:system-prompt}]
You are a security researcher tasked with evolving
vulnerability detectors for Joern. Joern uses Code Property
Graphs (CPG) to analyze codebases and find security
vulnerabilities. You will see a corpus of Python code in
{CORPUS_PATH}.

analyse_corpus() returns a compact summary (findings per
query) and writes detailed critic feedback to
`/repo/analysis_details.json`. Read that file for the
critic's qualitative assessment (coverage gaps, FP
patterns). Maximize real detections (true positives) while
minimizing false positives.

## Your objective
The only thing measured is detector quality. Run
`analyse_corpus()` to see how your detectors perform on the
corpus; when the results improve, call `submit(desc="...")`
to commit a new candidate and end your turn. That feedback
loop is the whole game -- your detectors are the deliverable.
Queries must be written in Joern's Scala-based query
language (`.sc` files); Python detector tools are also
allowed.

`analyse_corpus()` is a pure measurement step -- call it as
often as you like to test and refine; it does NOT commit.
The committed candidate persists to the next generation, so
iterate (analyse, read errors/critic, fix, re-analyse)
before you submit.

## How you get there is up to you
You're free to write whatever helps you do that well --
exploratory scripts, helper tools, notes, scratch files,
documentation -- anywhere under /repo. Nothing below is
mandatory ritual; they are aids that tend to pay off:
- Read /repo/NOTES.md -- your persistent lab notebook from
  prior generations (detectors that already work and
  shouldn't be broken, FP-prone patterns, ideas to try).
  Build on it instead of starting cold.
- Explore the corpus and skim /repo/docs/ before writing a
  detector, so your query matches patterns that are
  actually present.
- Jot down what each analyse_corpus() run taught you in
  NOTES.md while it's fresh -- which query files yield real
  true positives, which patterns were FP-prone, what to try
  next -- so the next generation inherits it.

Do these if they help; skip them if they don't. Spend your
effort where it moves detector quality.

## Environment
Your working directory is `/repo` with your mutable
detector code:
- `/repo/joern_analyser/` -- your working copy (changes
  here are evolved)
- `/repo/docs/` -- documentation for reference
- `/corpus/` -- read-only analysis corpus

## Available Tools
- `think()` -- reason through your approach
- `bash(command="...")` -- run shell commands in the
  container (write files, explore, grep). `/repo/` is your
  workspace, `/corpus/` is the corpus
- `delete_query(name="X.sc")` -- delete a detector you
  wrote (use to remove a query that won't compile)
- `analyse_corpus()` -- run detectors on corpus and
  measure quality (expensive, ~2 min); returns
  findings-per-query counts and critic feedback.
  Repeatable -- does NOT commit. Call it freely to test
  and iterate.
- `submit(desc="...")` -- commit the current detectors as
  a new generation and end your turn. Runs a fresh
  analysis, archives the candidate with its score, then
  ends the generation. Call this only when you are
  satisfied; you cannot edit afterwards.
- `ticket(action, id, note)` -- inspect and act on FP
  (false-positive) tickets the critic opens in
  /repo/tickets/. `ticket(action="list")` shows open
  tickets; `ticket(action="show", id=...)` prints one. To
  clear a ticket that really is a false positive, tighten
  its named query so it stops firing, then re-run
  analyse_corpus(). If you believe the finding is a REAL
  vulnerability, `ticket(action="dismiss", id=...,
  note="why it's real")` -- the critic re-judges with your
  argument and only closes the ticket if convinced
  (limited attempts).
\end{lstlisting}

\subsection{Generation Prompt}
\label{sec:prompt-generation}

Each generation receives this prompt, which defines the fitness function, scoring insights, valid vulnerability types, the parent workspace, and the detector-writing workflow with syntax constraints.

\begin{lstlisting}[style=python,language={},basicstyle=\ttfamily\scriptsize,caption={Generation prompt (verbatim, template variables in braces).},label={lst:generation-prompt}]
## Generation {gen} of {total}

MESSAGE BUDGET: {max_messages} messages. Focus.

{universe_line}Workspace: branched from {parent_id}

## Goal: Maximize Fitness

Fitness = sqrt(precision x type_coverage)

Where:
- Precision = mean of per-group precision scores. Each
  query group (findings from one .sc file) is independently
  judged by an LLM that samples a few findings and decides
  if they are real vulnerabilities.
- Type coverage = fraction of the target vulnerability
  classes that your detectors actually exercise (emit at
  least one finding for). This is a density-independent
  recall proxy -- covering MORE distinct vulnerability
  classes raises it, regardless of how many files each
  fires on.

### Key Scoring Insights
- Both precision AND type coverage must be strong. A
  firehose that sprays one class across every file but is
  mostly FP scores poorly (low precision). A perfect
  detector that only covers 1 vulnerability class also
  scores poorly (low coverage).
- Groups are evaluated independently. A group scoring 0.0
  drags down mean precision.
- Better to have 3 groups at precision=1.0 covering 3
  different vulnerability classes than 5 groups, all of the
  same class, where 2 score 0.0.
- The judge samples ~5-8 findings per group. Very small
  groups (<5 findings) are judged in full.
- Type coverage rewards detecting MANY DISTINCT
  vulnerability classes, not many files. To raise it, add
  detectors for classes that currently produce no findings
  -- spraying more findings of a class you already cover
  does not help coverage.
- If findings haven't changed between runs, scores are
  cached (deterministic). Only new/changed findings get
  re-judged.

## Valid Vulnerability Types

Your query filename MUST use one of these types as prefix
(e.g., `SQL_INJECTION__HIGH__desc.sc`).
Findings with unrecognized types are classified as UNKNOWN
and automatically score precision=0.0:

SQL_INJECTION, COMMAND_INJECTION, XSS, PATH_TRAVERSAL,
CODE_INJECTION, LDAP_INJECTION, XXE, DESERIALIZATION,
SSTI, IDOR, BROKEN_ACCESS_CONTROL,
BROKEN_AUTHENTICATION, INFORMATION_DISCLOSURE,
CRLF_INJECTION, LOG_INJECTION, SSRF,
WEAK_CRYPTOGRAPHY, WEAK_ENTROPY, WEAK_RANDOMNESS,
XPATH_INJECTION, INSECURE_COOKIE, OPEN_REDIRECT,
SESSION_FIXATION, TRUST_BOUNDARY_VIOLATION

## Current State
{metrics_display}

## Writing Detectors

CRITICAL: Only write a detector if you understand taint
analysis. Do NOT write bare `cpg.call().l.foreach`
queries -- they don't find real vulnerabilities. Use
`reachableByFlows(SOURCE_PATTERNS)` for taint tracking.

Steps:

1. Analyze corpus samples to identify:
   - What sink functions should be detected?
   - What argument positions contain user input?
   - What taint sources are relevant? (request
     parameters, uploaded files, etc.)
   - What vulnerability name to use?

2. Write a taint-tracking query. Example (SQL Injection):
   cpg.call("execute|executeQuery|executeUpdate")
     .argument(0)
     .reachableByFlows(
       cpg.method.name(".*request.*|.*getParam.*")
       .parameter)
     .l
     .foreach { node =>
       val lineNum = node.lineNumber
         .getOrElse(-1).toString
       println(s"${node.file}|||${lineNum}|||"
         + s"SQL Injection|||${node.code}")
     }
   - Write to /repo/joern_analyser/analyses/custom/
     queries/TYPE__SEVERITY__description.sc
   - Use bash heredocs: bash(command="cat >
     /repo/.../X.sc << 'EOF'\n...\nEOF")
   - Use single-quoted EOF delimiter to avoid shell
     interpolation of $ in Scala

3. Analyze with analyse_corpus() to evaluate fitness

## Workflow (MANDATORY - FOLLOW IN THIS ORDER)

PHASE 0: READ DOCUMENTATION FIRST
- docs/fundamentals/cpg.md -- understand Code Property
  Graph
- docs/query-language/cpgql-reference.md -- how to write
  queries
- docs/detectors/sql-injection.md -- example detector
  patterns
- KEY RULE: DO NOT include `importCpg` in your query --
  framework adds it
- Check existing detectors (bash grep) to see working
  patterns

### CRITICAL: Scala/CPGQL Syntax Errors to Avoid

DO NOT use `reachableBy(sources:_*)` -- this DOES NOT
COMPILE in Joern 4.x.

The `:_*` splat annotation is invalid for
`reachableBy`/`reachableByFlows` parameters.

WRONG (will fail with "Sequence argument type annotation
`*` cannot be used here"):
  val sources = cpg.method.name(".*param.*").parameter
  cpg.call("exec").argument(0)
    .filter(c => c.argument
    .reachableBy(sources:_*).nonEmpty)

CORRECT -- pass the traversal directly:
  cpg.call("exec").argument(0)
    .reachableByFlows(
      cpg.method.name(".*param.*").parameter)
    .l
    .foreach { flow => ... }

Always use the `reachableByFlows` taint-tracking pattern
-- do NOT invent new taint-tracking patterns or write
bare `cpg.call` queries.

PHASE 1: EXPLORE CORPUS (now that you understand Joern)
{corpus_hint}

Read actual files from the corpus. Grep for patterns.
Identify what vulnerabilities are actually present.

PHASE 2: IMPLEMENT WITH JOERN (only after understanding
Joern + corpus)
Design and implement detectors using proper Scala/CPGQL
syntax from the docs.
Test with analyse_corpus() -- each run automatically
commits the current state as a new generation and seeds
the next.
\end{lstlisting}

\subsection{Critic Feedback Template}
\label{sec:prompt-critic}

After each evaluation, the critic returns a structured summary injected into the agent's context to guide the next mutation.

\begin{lstlisting}[style=python,language={},basicstyle=\ttfamily\scriptsize,caption={Critic feedback template (verbatim).},label={lst:critic-template}]
  ESTIMATED PRECISION:  {precision:.2%}
    ({tp_verified} of {total_findings} sampled findings
     verified genuine)
  COVERAGE SCORE:       {coverage:.2%}
    (detector found patterns in ~{coverage_pct}% of
     sampled files)
  CRITIC FITNESS:       {fitness:.2%}
    (goal: approach 1.0)
\end{lstlisting}

\subsection{Critic Precision Judge}
\label{sec:prompt-judge}

The precision half of the fitness signal is produced by an \ac{LLM} judge. For each detector group (the findings emitted by one \texttt{.sc} query), the critic samples a few representative findings spanning distinct files and asks the model to count how many are genuinely exploitable; the group's precision is the fraction judged real, and overall precision is the mean across groups. Sampling scales with group size (all findings if $\leq 5$, up to $8$ for medium groups, $5$ for large groups), and unchanged findings are cached to keep scoring deterministic across runs.

\begin{lstlisting}[style=python,language={},basicstyle=\ttfamily\scriptsize,caption={Critic precision-judge prompt (verbatim; \texttt{\{n\_sampled\}}, \texttt{\{n\_total\}}, \texttt{\{vuln\_type\}}, and \texttt{\{findings\_json\}} are filled per group at runtime).},label={lst:judge-prompt}]
You are a security expert. Below are {n_sampled}
representative findings from a group of {n_total}
detector findings of type {vuln_type}.
For EACH, decide if it is a REAL exploitable
vulnerability (not just a code smell or theoretical
concern). Be strict: a finding is real only if
untrusted input can reach a dangerous sink without
proper sanitization.

{findings_json}

Respond with JSON:
{"real_count": <int 0-{n_sampled}>, "reason": "..."}
\end{lstlisting}

\end{document}